# Estimating Sparse Sources from Data Mixtures using Maxima in Phase Space Plots

Malcolm Woolfson

*Department of Electrical and Electronic Engineering, Faculty of Engineering, University of Nottingham, Nottingham. NG7 2RD. England*

email: malcolm.woolfson@outlook.com

**Abstract**

In Blind Source Separation (BSS), one estimates sources from data mixtures where the mixing coefficients are unknown. In the particular case of Sparse Component Analysis (SCA), each underlying source exists for only a finite amount of time when other sources are negligible. In this paper, one approach to SCA is presented where the data are represented using phase space analysis and one estimates the main source from the maximum in the phase plot. Deflation is used to estimate the other sources. The proposed method is tested on simulated data and experimental ECG data taken from an expectant mother. It is shown that, in most cases, the performance of the proposed method is comparable to that of Principal Component Analysis (PCA) and FastICA for clean data. In the case of noisy data, PCA is found to be more robust for higher noise levels. For situations where the sources have coincident peaks, the method breaks down as expected, as the maximum in the phase plot does not correspond to an individual source.

## 1. Introduction

One general problem in signal processing is the extraction of individual source signals $\{s_j[n]\}$ from measurements $\{z_i[n]\}$ that are a linear combination of these sources:

$$z_i[n] = \sum_{j=1}^{N} A_{ij} s_j[n] \tag{1}$$

where $i = 1, 2, \ldots, S$ with $S$ being the number of sets of measurement data, $\{A_{ij}\}$ are the mixing coefficients, and there are $N$ underlying sources. In the case where both the sources and mixing coefficients are unknown, then this problem comes under the heading of Blind Source

Separation (BSS). In this paper, we will assume that the number of sources equals the number of measurements, $S = N$.

There are many applications in this area, for example the analysis of EPR data [1], NMR data [2], fetal ECG monitoring [3] and gene mapping [4].

One of the simplest approaches to this problem is to use Principal Component Analysis (PCA) [5] where one is finding a combination of data inputs that maximises the variance; this can be found using an eigenanalysis of the data covariance matrix. When applying this method, it is assumed that the underlying sources are uncorrelated but this restriction is not imposed on the higher order cross-moments that are required to make the estimated sources independent. Methods that are more sophisticated concentrate on zeroing the fourth cross-moments [6], others on minimising the Gaussian nature of the estimated sources [7],[8]. These methods involve applying PCA as a pre-processing step in order to whiten the data.

The above-mentioned methods are appropriate for general sources which have non-zero values for all times, and which are not necessarily sparse. Various workers, for example in References [9-27], have looked at the general problem of extracting sparse sources. The main problem in Sparse Component Analysis (SCA) is to identify segments of the data where one source exists on its own, or at the very least dominates the other sources, and use this information to estimate that source and then use deflation to estimate the other sources.

One approach to estimating sparse sources in data mixtures is phase space analysis where, for $N$ data inputs, the data are plotted against each other in $N$-dimensional space. This approach has been used in SCA, where clustering is applied to the detection of straight lines in phase space representing individual sources. One approach, e.g. [16], is to apply clustering of data points to determine the dominant directions in phase space corresponding to the sources and then one uses matrix inversion to estimate the individual sources. Another approach [25] using phase space analysis is to determine these directions using a sorting procedure and then apply deflation to determine the underlying sources.

In this paper, we will look at the application of PCA to mixtures of sparse sources and investigate a relatively simple post-processing step, using phase space analysis, that can improve the estimates of the underlying sources.

## 2. The Maximum Method

In some approaches to Blind Source Separation, it is convenient to represent the data in phase space. This is easiest to demonstrate in two dimensions.

Suppose that we have two underlying sources that are uncorrelated as shown in Figure 1:

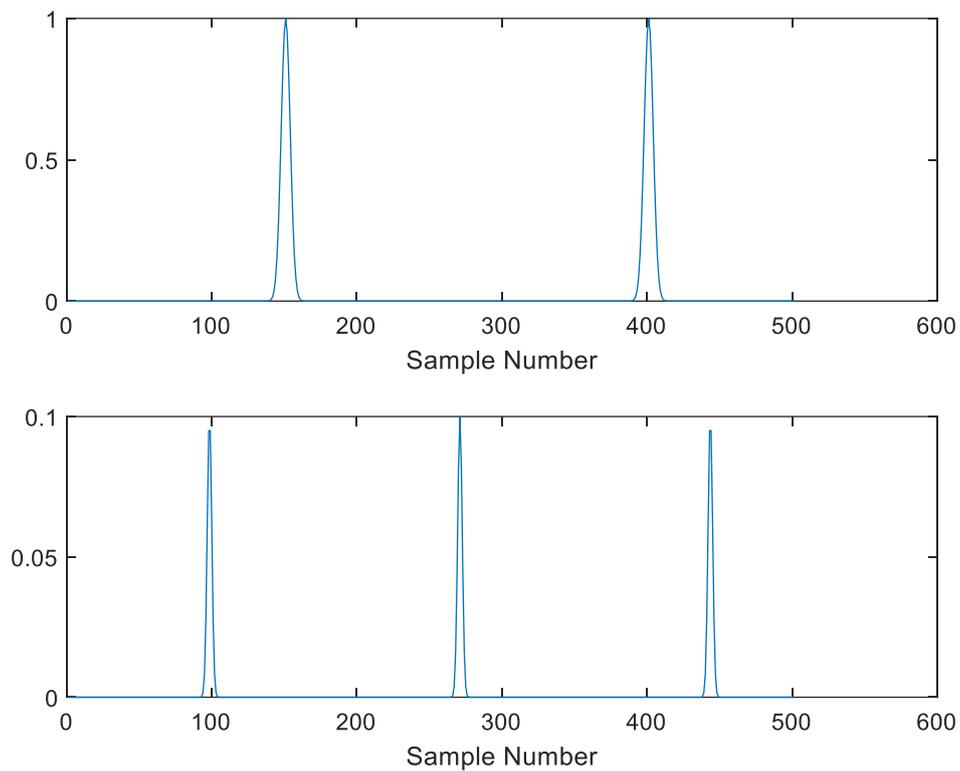

*Figure 1 – Two non-overlapping sources*

Now suppose that these data are mixed with the following mixing matrix:

$$\boldsymbol{A} = \begin{pmatrix} 1.3 & 2 \\ 1 & 3 \end{pmatrix} \qquad (2)$$

The mixed data are shown in the Figure 2 below:

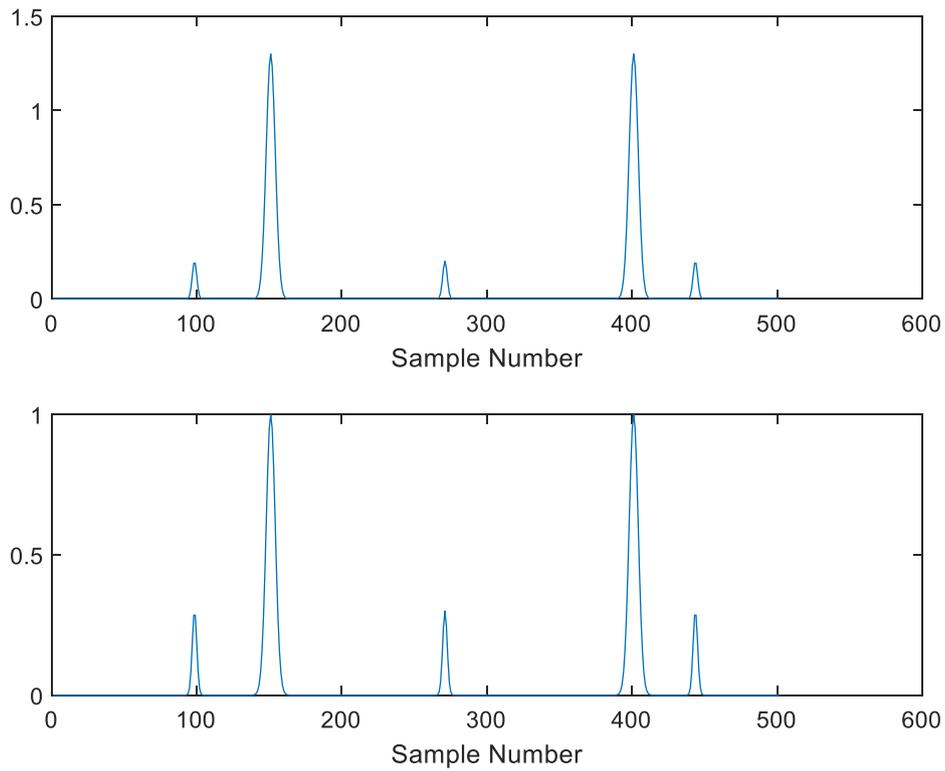

*Figure 2 – Mixed data applying the mixing matrix (2)*

In order to plot this data in phase space, we take $z_1$ as the x-variable and $z_2$- as the y variable and plot $(z_1, z_2)$ pairs for each sample point and then join up the points to produce the following phase plot shown in blue – ignore the black line for the moment:

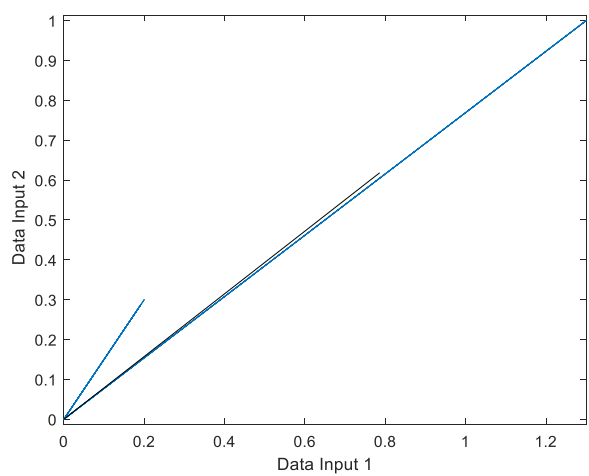

*Figure 3 – Phase plot (blue lines) of mixed data in Figure 2. Black Line is principal direction found using PCA*

It can now be seen that there are two dominant directions in phase space (shown in blue), each corresponding to an individual source.

When PCA is applied, the direction in phase space is chosen such that the projection of the data onto this direction has the maximum variance. In this paper, we do not carry out normalisation of the input data when using PCA as this always results in fixed eigenvectors for all mixtures. Centring of the data is also not used in this example for reasons explained later. If PCA is applied to the above data, the direction of maximum variance is shown as the black line in Figure 3. Note that this line does not correspond to the direction of one of the sources, although it is close to one of these lines. PCA processes the whole data and the direction where the projections have the maximum variance does not always correspond to the direction of a particular source; this results in the estimate of one source being contaminated by the other. The estimates of the sources when applying PCA are shown in the Figure 4 below:

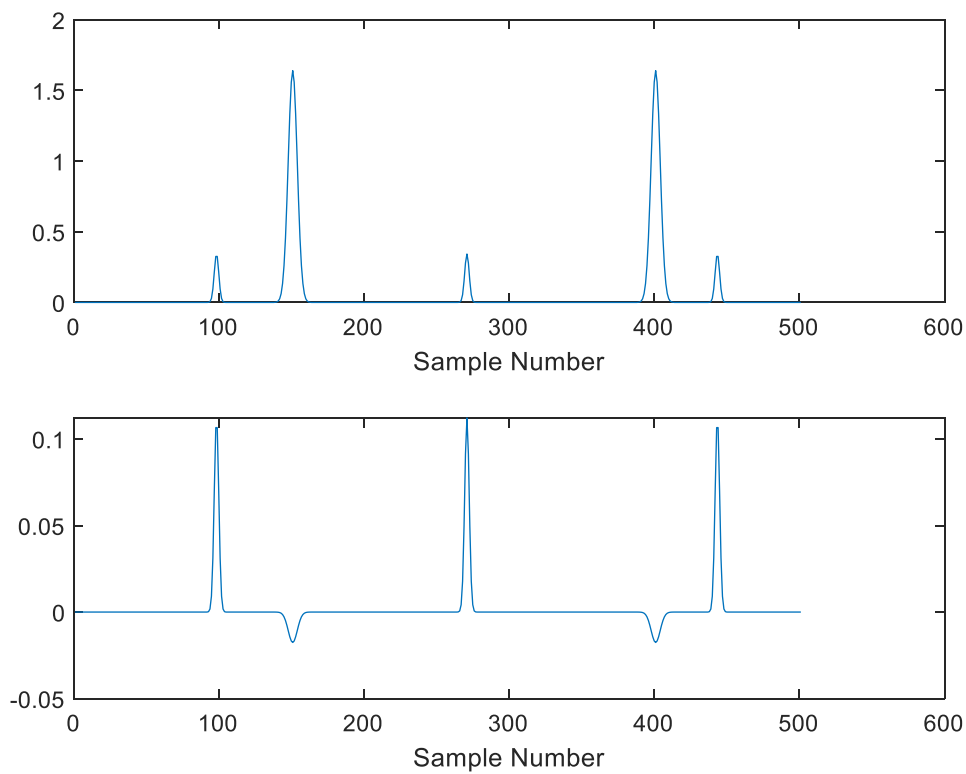

*Figure 4 – Source Estimates using PCA*

A potentially better solution is if the two directions corresponding to individual sources can be found directly from the phase plot, Figure 3. One simplified way of doing this is to determine the point of maximum distance of the trajectory in the phase plot from the origin and use the vector joining the origin to this point of maximum direction to determine the principal source.

Suppose that we have an $N$-dimensional example, where the data inputs at each sample $n$ is represented by the following points in the phase plot

$$\mathbf{z}[n] = ((z_1[n], z_2[n], \ldots, z_N[n]))'$$

($n$=1,2,…,$M$) (3)

The distance of each point in the phase plot from the origin is given by

$$r[n] = |\mathbf{z}[n]| = \sqrt{z_1^2[n] + z_2^2[n] + \cdots + z_N^2[n]}$$

($n$=1,2,…,$M$) (4)

The sample point, $n_{max}$ corresponding to the most significant source is given by the value of $n$ such that $r[n]$ is a maximum. The direction in the phase plot corresponding to the most significant source is then taken as the unit vector $\hat{\mathbf{z}}[n_{max}]$. We will refer to this method as the Maximum method.

Using deflation the contribution of the principal source to the data can be subtracted out and the remaining sources can then be found using an iterative process, which will be described in Section (3) below.

The potential advantage of using the Maximum method compared to using PCA is that the actual direction in phase space corresponding to each source can be found and this direction is not influenced by other sources as long as they are uncorrelated. As we will see, a potential disadvantage of the proposed method is that it is more sensitive than other methods to situations where there is coincidence between the peaks of different sources.

We can now use this direction of the Maximum to estimate the most prominent source. A detailed analysis is given in Reference [25]; just a summary is given here.

The vector, $z$, representing the phase plot can be written as a linear combination of the vectors representing the sources, $\mathbf{R}_i$, as follows:

$$\mathbf{z} = \sum_{i=1}^{N} s_i \mathbf{R}_i \tag{5}$$

Where we have dropped the sample dependence [$n$]. There are assumed to be $N$ underlying sources $\{s_i\}$, i=1,2,…,$N$, under the following assumptions:

(1)  The number of mixed data signals equals the number of sources

(2)  For each source there is a specific time slot such that all the other sources are zero.

In this analysis, lower case symbols refer to variables that are a function of time, whilst upper case symbols refer to time-independent variables. For notational convenience, the time-dependencies are suppressed. Note that this notation means that some vectors will be represented by upper case variables, but the vector nature of such variables will be indicated.

Suppose that the direction of the Maximum in phase space is given by the vector

$$\mathbf{R}_p = \mathbf{z}[n_{max}]$$

Let the unit vector in this direction be $\widehat{\mathbf{R}}_p$.

An estimate of this source signal, $\tilde{s}_p$, is obtained from the phase plot by finding the component of the phase plot data in this direction:

$$\tilde{s}_p = \widehat{\mathbf{R}}_p \cdot \mathbf{z} \tag{6}$$

Substituting for $\mathbf{z}$ from Equation (5):

$$\tilde{s}_p = \widehat{\mathbf{R}}_p \cdot \sum_{i=1}^{N} \mathbf{R}_i s_i \tag{7}$$

Writing

$$\mathbf{R}_i = R_i \widehat{\mathbf{R}}_i \tag{8}$$

Equation (7) can be rewritten as

$$\tilde{s}_p = \widehat{\mathbf{R}}_p \cdot \sum_{i=1}^{N} \widehat{\mathbf{R}}_i R_i s_i \tag{9}$$

After further rearranging Equation (9) can be written as:

$$\tilde{s}_p = \sum_{i=1}^{N} s_i R_i (\widehat{\boldsymbol{R}}_p \cdot \widehat{\boldsymbol{R}}_i) \tag{10}$$

Separating out the term $i = p$ in the summation (10):

$$\tilde{s}_p = s_p R_p + \sum_{i=1(i\neq p)}^{N} s_i R_i (\widehat{\boldsymbol{R}}_p \cdot \widehat{\boldsymbol{R}}_i) \tag{11}$$

In the general case where $(\widehat{\boldsymbol{R}}_p \cdot \widehat{\boldsymbol{R}}_i) \neq 0$, $\tilde{s}_p$ will depend on both $s_p$ and the other sources through the second term in (11).

## 3. The Method of Deflation

In Section 2, the Maximum method was described to determine the most prominent source. In order to determine the other sources present, we use the deflation method. Deflation is a general method that can be used to subtract out individual sources that have just been estimated to determine other sources in the mixture. Details have been given in [25] for this method and this is repeated here for completeness.

In the phase space plot the estimate of the dominant source, $\tilde{s}_p$ is represented by a vector

$$\tilde{\boldsymbol{s}}_p = \tilde{s}_p \widehat{\boldsymbol{R}}_P \tag{12}$$

Substituting for $\tilde{s}_p$ from Equation (11) into (12):

$$\tilde{\boldsymbol{s}}_p = s_p R_p \widehat{\boldsymbol{R}}_P + \widehat{\boldsymbol{R}}_P \sum_{i=1(i\neq p)}^{N} s_i R_i (\widehat{\boldsymbol{R}}_p \cdot \widehat{\boldsymbol{R}}_i) \tag{13}$$

We now subtract this vector from the original vector **z:**

$$\boldsymbol{z}' = \boldsymbol{z} - \tilde{\boldsymbol{s}}_p \tag{14}$$

as a first step to find the other sources.

Using Equations (5), (13) and (14), we can express $\boldsymbol{z}'$ as follows:

$$\mathbf{z}' = \sum_{i=1}^{N} s_i R_i \widehat{\mathbf{R}}_i - s_p R_p \widehat{\mathbf{R}}_p - \widehat{\mathbf{R}}_p \sum_{i=1(i\neq p)}^{N} s_i R_i (\widehat{\mathbf{R}}_p \cdot \widehat{\mathbf{R}}_i) \tag{15}$$

Combining the first two terms into one:

$$\mathbf{z}' = \sum_{i=1(i\neq p)}^{N} s_i R_i \widehat{\mathbf{R}}_i - \widehat{\mathbf{R}}_p \sum_{i=1(i\neq p)}^{N} s_i R_i (\widehat{\mathbf{R}}_p \cdot \widehat{\mathbf{R}}_i) \tag{16}$$

which can be further rewritten as:

$$\mathbf{z}' = \sum_{i=1(i\neq p)}^{N} s_i R_i [\widehat{\mathbf{R}}_i - \widehat{\mathbf{R}}_p (\widehat{\mathbf{R}}_p \cdot \widehat{\mathbf{R}}_i)] \tag{17}$$

Equation (17) can be written in the more compact form:

$$\mathbf{z}' = \sum_{i=1(i\neq p)}^{N} s_i \mathbf{R}'_i \tag{18}$$

where

$$\mathbf{R}'_i = R_i [\widehat{\mathbf{R}}_i - \widehat{\mathbf{R}}_p (\widehat{\mathbf{R}}_p \cdot \widehat{\mathbf{R}}_i)] \tag{19}$$

In this case the time varying vector $\mathbf{z}'$ in Equation (18) does not depend on the source $s_p$ which has been subtracted out of the calculation; this vector depends on the other $N$-1 sources that have not yet been estimated. The above procedure is iterated, applying the Maximum method to $\mathbf{z}'$ until there are no components left. Our final estimates for the sources are $\{\tilde{s}_p\}$ $p = 1,2,\ldots,N$.

If we apply the Maximum method to the data mixtures shown in Figure 2, we obtain the estimates of the two sources shown in Figure 5 below:

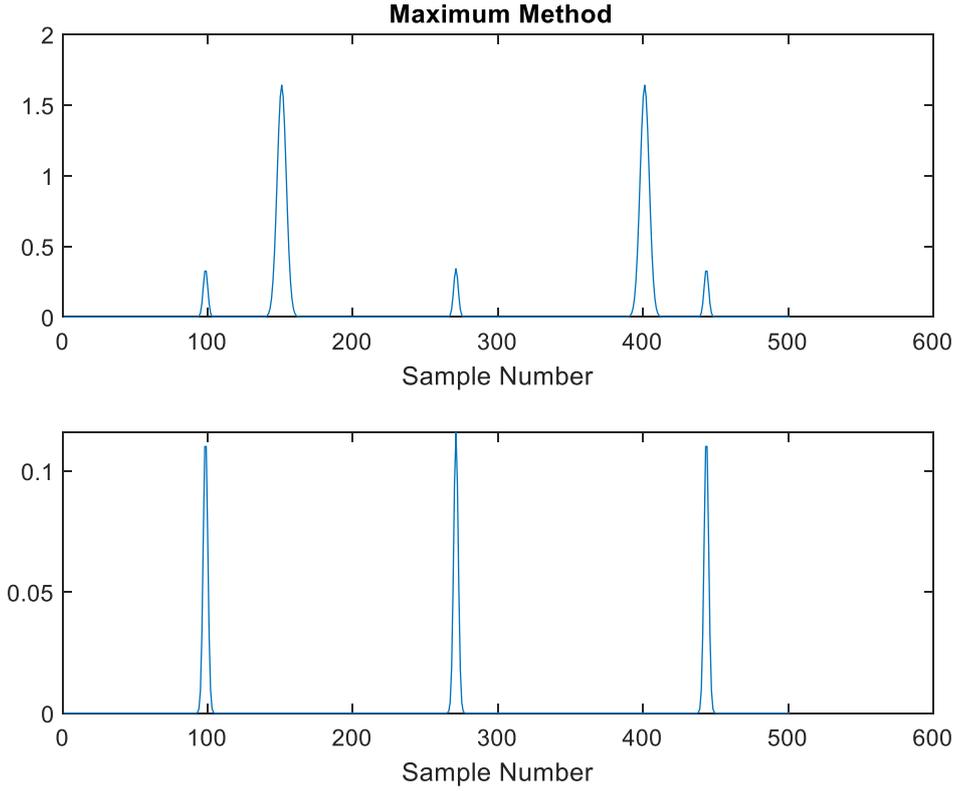

*Figure 5 – Source estimates using the maximum method.*

It can be seen that the estimate of one of the sources is contaminated by the other (top graph in Figure 5) whilst the second source is estimated to within a scaling constant (bottom graph in Figure 5). To explain why the first source estimate is contaminated by the second, let us look specifically at the case of two sources, $N = 2$; from Equation (11) the estimate of Source 1 becomes

$$\tilde{s}_1 = \sum_{i=1}^{2} s_i R_i (\widehat{\boldsymbol{R}}_1 . \widehat{\boldsymbol{R}}_i)$$
$$= s_1 R_1 + s_2 R_2 (\widehat{\boldsymbol{R}}_1 . \widehat{\boldsymbol{R}}_2) \tag{20}$$

After subtraction of this source, the vector representing the phase plot of the subtracted data is given by (18) with $N = 2$:

$$\boldsymbol{z}' = \sum_{i=1(i \neq 1)}^{2} s_i \boldsymbol{R}'_i = s_2 \boldsymbol{R}'_2 \tag{21}$$

where

$$\boldsymbol{R}'_2 = R_2\left[\widehat{\boldsymbol{R}}_2 - \widehat{\boldsymbol{R}}_1(\widehat{\boldsymbol{R}}_1.\widehat{\boldsymbol{R}}_2)\right] \tag{22}$$

and an estimate of source $s_2$ can be determined from

$$\widetilde{s_2} = \boldsymbol{z}'.\widehat{\boldsymbol{R}'}_2 = s_2 R_2' \tag{23}$$

It can be seen that the estimate of the first source (20) has a contaminating term $s_2 R_2(\widehat{\boldsymbol{R}}_1.\widehat{\boldsymbol{R}}_2)$ from the second source. The second source (23) is estimated to within a scaling constant; the results of the calculation are in agreement with the theory. Looking at the phase plot of the data in Figure 3, it can be seen that the directions corresponding to the two sources are not orthogonal so that $\widehat{\boldsymbol{R}}_1.\widehat{\boldsymbol{R}}_2 \neq 0$. Now the original sources are uncorrelated. However, if we apply whitening to the data then, for uncorrelated sources, the directions in phase space corresponding to the sources are orthogonal and $\widehat{\boldsymbol{R}}_1.\widehat{\boldsymbol{R}}_2 = 0$. Whitening involves processing **z** using a linear transform to produce vectors that are orthonormal. One such method is PCA where one normalises the eigenvectors. Another method that has been applied in phase space methods [25] is the Gram-Schmidt method with normalisation of vectors to produce an orthonormal set of data; if this method is applied to the data **z** to produce the whitened components **e** = ($e_1$,$e_2$), then the phase plot is displayed in Figure 6 below:

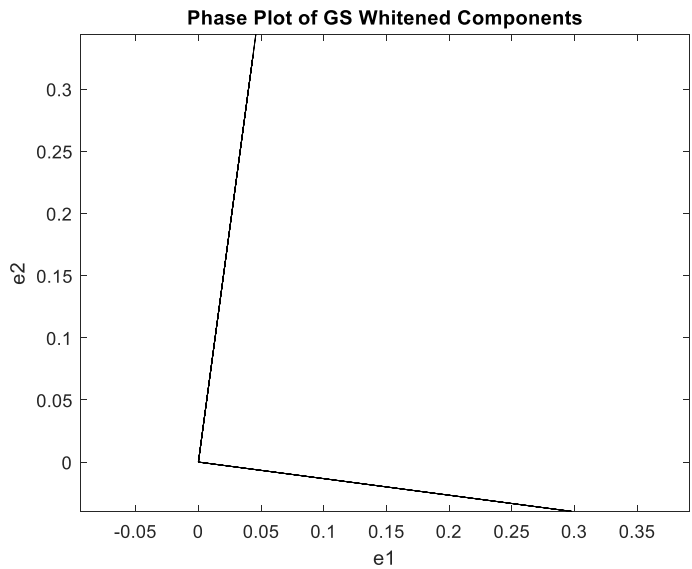

*Figure 6 – Phase plot of whitened data*

and the directions corresponding to the two sources are now orthogonal with $\widehat{R}_1 . \widehat{R}_2 = 0$; substituting this relation into Equations (20), (22) and (23), the estimates for the two sources become:

$$\tilde{s}_1 = s_1 R_1 \qquad (24)$$

and

$$\tilde{s}_2 = s_2 R_2 \qquad (25)$$

From Equations (24) and (25) we can then see that both sources can now be estimated to within a scaling constant and this is verified in Figure 7 where the Maximum method is applied to the pre-whitened data **e** and the estimated sources are plotted out.

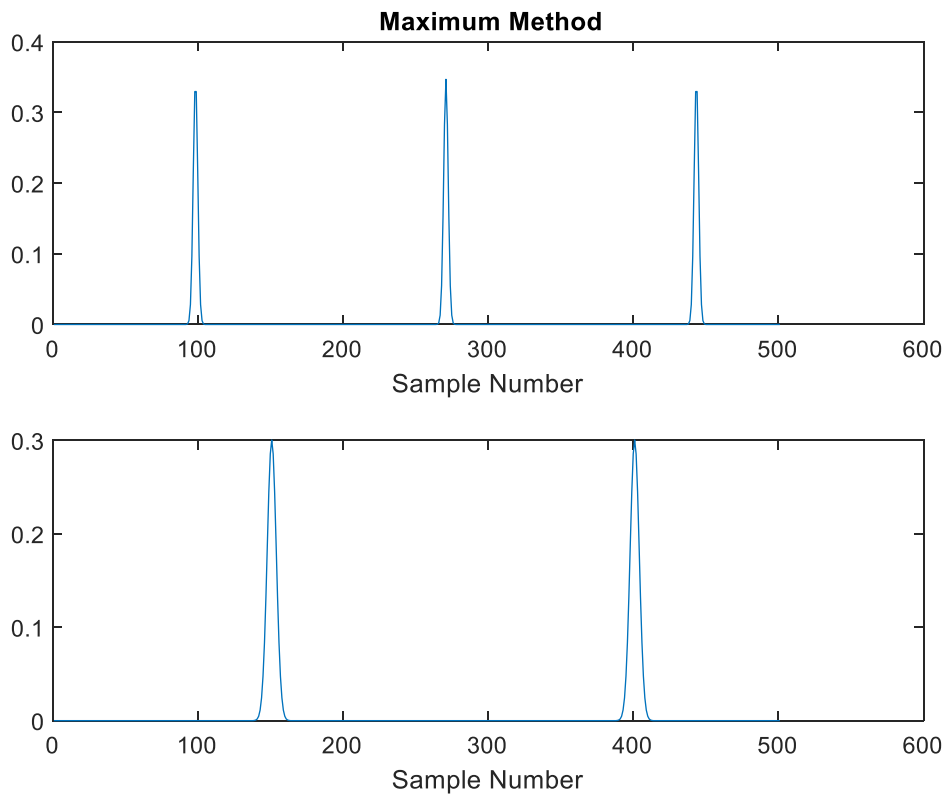

*Figure 7 – Sources estimated using the maximum method applied to whitened data*

To summarise, for this example, the Maximum method works better than PCA for mixtures of uncorrelated sparse sources as it is able to determine the direction of each source in phase space and hence estimate each source separately. PCA processes the whole data and the direction where the projections have the maximum variance does not necessarily correspond to the direction of a particular source.

## 4. Results

In this section, the Maximum method is applied to a variety of mixtures of sources and a comparison will be made with PCA and Fast ICA to determine if there are any improvements using the proposed method.

### 4.1 Two Pure Sources

The simplest case to look at is if the two inputs are pure unmixed sources. Any method of BSS should be able to extract the original sources.

As an example, the phase plot for two sources in Figure 1 is shown in Figure 8 below; it can be seen that the directions corresponding to the two sources are orthogonal reflecting that these sources are uncorrelated. It should be noted that, in the particular case of two pure sources, there is no need for pre-whitening to produce a phase plot with orthogonal directions; however, we will still apply whitening for consistency with other calculations that will be presented later.

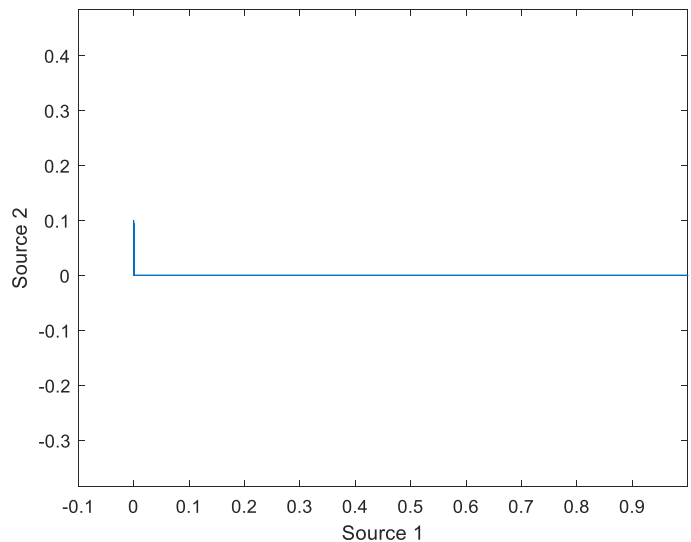

*Figure 8 – Phase plot of unmixed sources*

If we apply the Maximum method with pre-whitening using the Gram-Schmidt method, then the estimated sources are produced to within a scaling constant as shown in Figure 9 below:

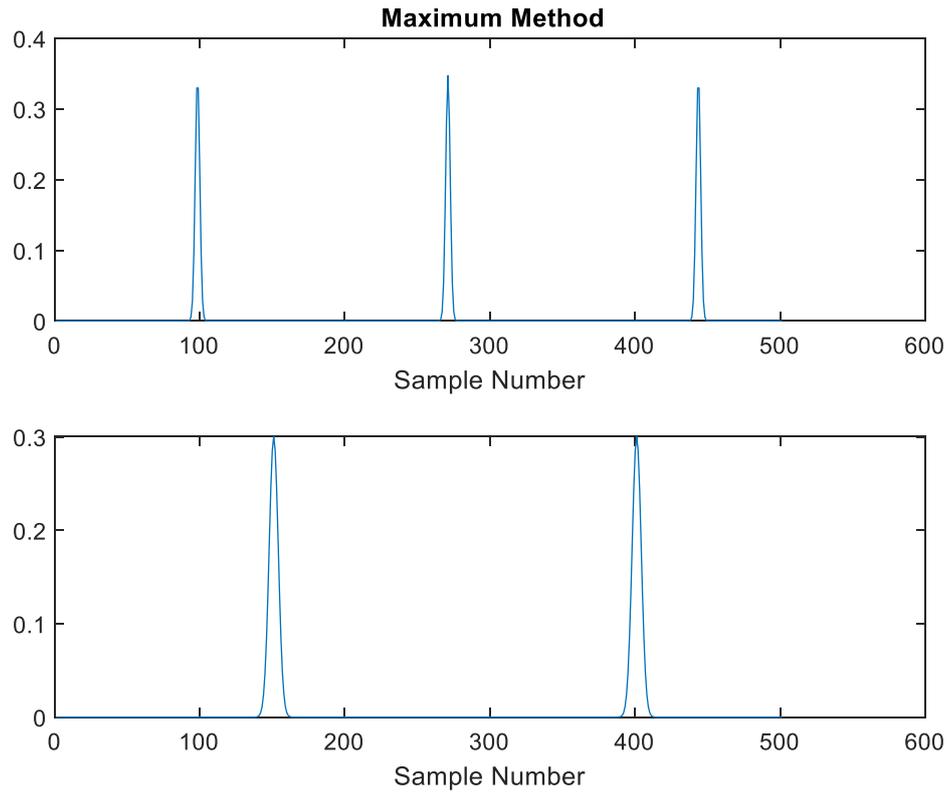

*Figure 9 – Sources estimated using the Maximum method*

If we apply PCA <u>without</u> subtracting the mean of the data, then the sources are also extracted to within a scaling constant as shown in Figure 10:

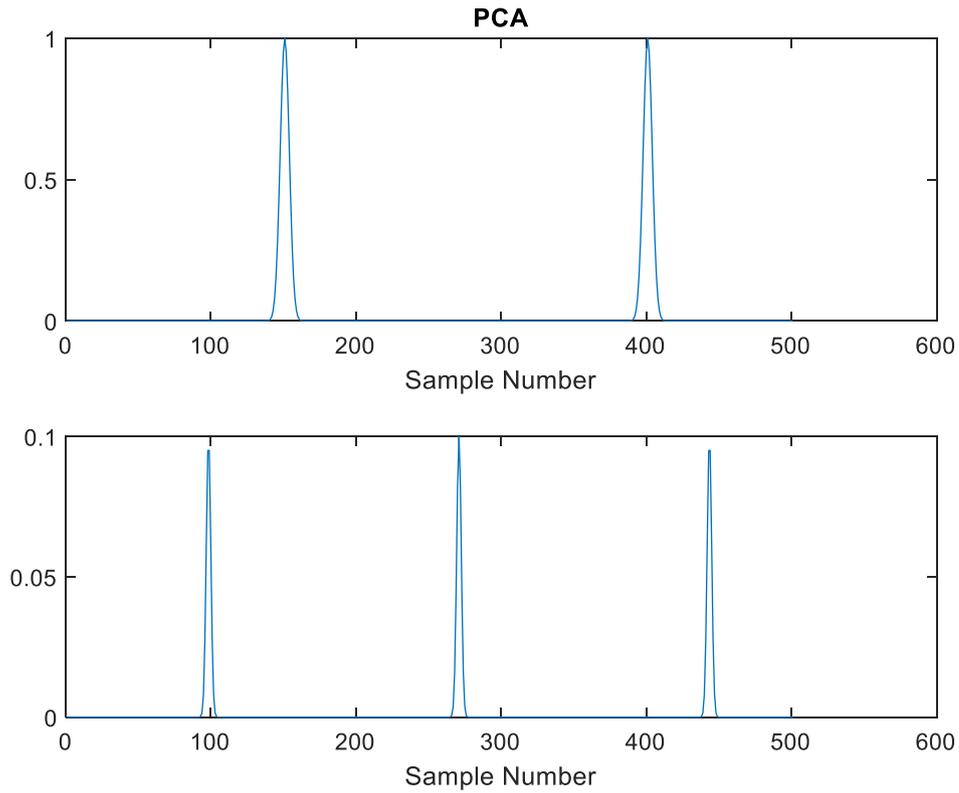

*Figure 10 – Sources estimated using PCA*

If we apply centring, that is subtract the means off each data input and repeat the above calculations, then we obtain the following estimated sources when using the Maximum method (Figure 11) and PCA (Figure 12).

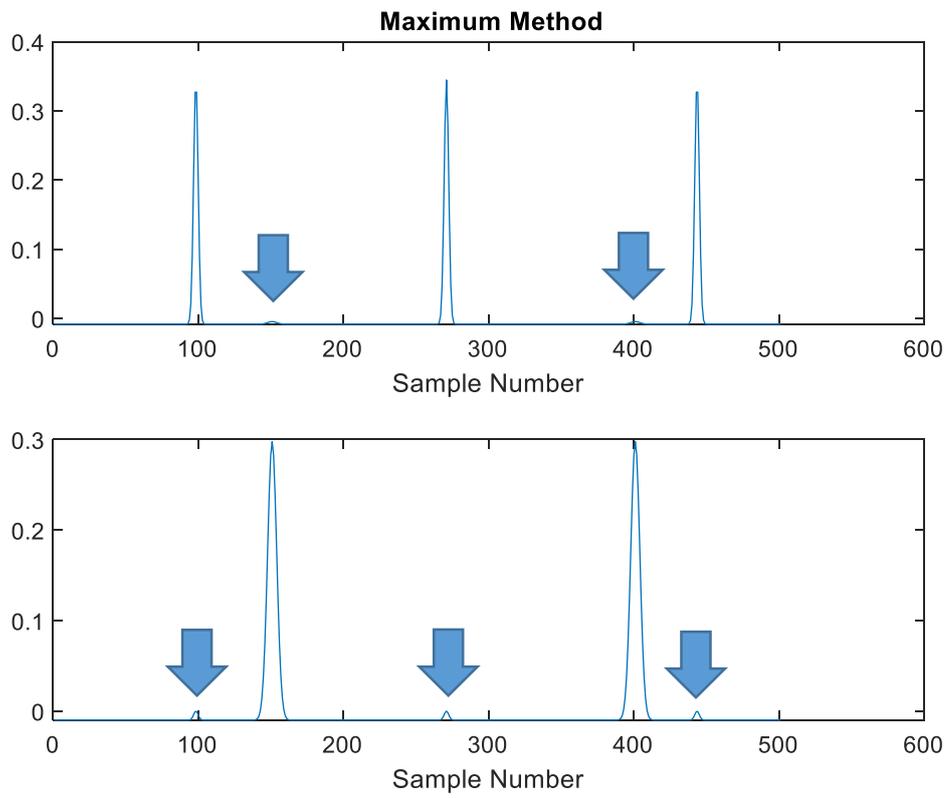

*Figure 11 – Sources Estimated using the Maximum method when centring is applied to the input data. Arrows indicate contamination of one source estimate by the other source.*

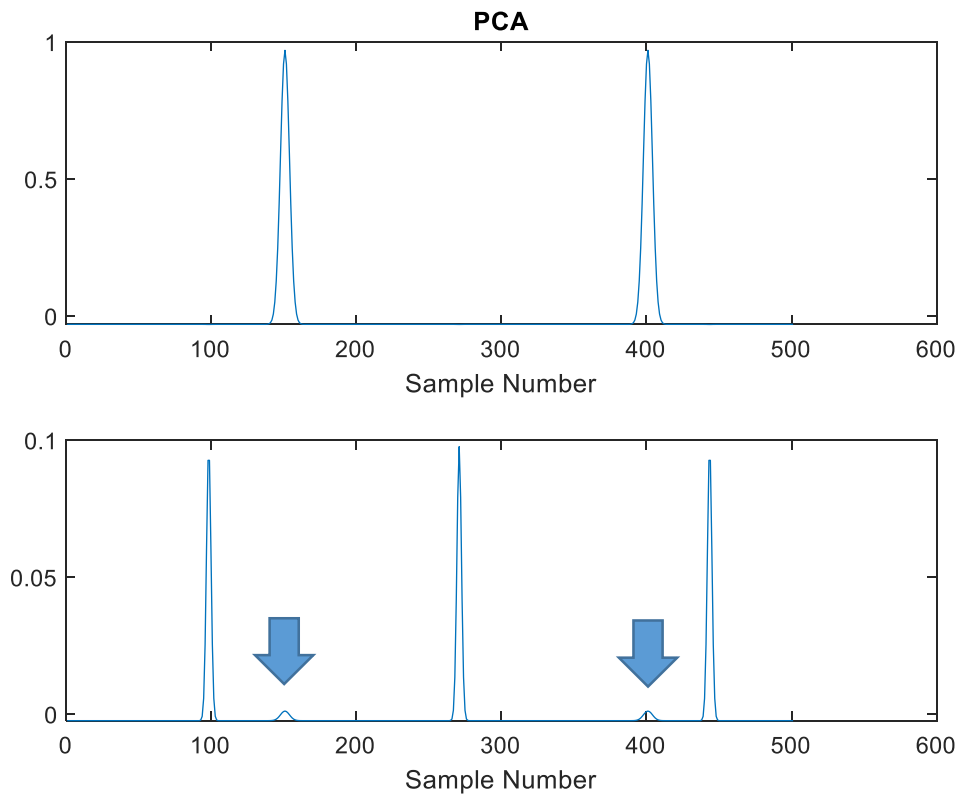

*Figure 12 – Sources estimated using PCA when centring is applied to the input data. Arrows indicate contamination of one source estimate by the other source.*

In the Maximum method (Figure 11), each estimated source is contaminated by the other source (indicated by the arrows). When PCA is applied (Figure 12), then one of the sources is contaminated (indicated by the arrows), but the other estimate is almost pure. Therefore, both methods are sensitive to centring. To see why, if we subtract the mean of each source and calculate the dot (scalar) product between the resulting sources, we obtain a non-zero value; this shows that the cross-correlation coefficient is no longer zero and the directions in phase space corresponding to the sources are no longer orthogonal. The baselines of the data, which were zero are now non-zero and these correlate with each other. The shifting of the baseline is clearly important: in the first set of calculations, it was assumed that the baseline is zero, but this will not always be the case. The baseline value needs to be considered before applying PCA or indeed any other method of BSS to ensure that the directions in phase space corresponding to the two sources are as close to orthogonal as possible.

We now compare the robustness to noise of the PCA and Maximum methods. We will be using the unmixed sources in Figure 1 as our "clean" data. We have seen, for this particular set of data, that both PCA and the Maximum method are able to extract the sources as long as centring is not used. Hence, by adding noise and comparing the robustness to this noise we are solely looking at differences between the two methods in coping with noise and not comparing different performances in extracting the sources themselves.

We will look at the rms errors in the source estimation as a function of sample number for various noise levels. In Section 4 in [25] it is explained in detail how estimated sources are associated with actual sources and the normalised rms error calculated. Firstly, the estimated sources and clean sources are normalised to unit magnitude to take into account different scaling of the clean sources and the estimates. The normalised sources and normalised estimates are compared pairwise and the associations with the highest correlation function between source and estimate are made. When comparing the robustness to noise, we compare the rms errors to the normalised <u>clean</u> sources to assess how significant the errors are compared to the original source. The rms errors are computed for 1000 Monte Carlo runs.

Note from Figure 1 that, for the clean sources, the source with the smallest amplitude is Source 2 with amplitude 0.1; we will present simulations for various noise levels that are a percentage of this amplitude. Gaussian noise will be assumed.

Whitening using the Gram-Schmidt method will be used, taking Source 1 as the first vector. This is unnecessary for pure sources, as the directions in phase space corresponding to the two sources will be orthogonal. However, for mixed sources, whitening is needed to ensure that the directions in phase space corresponding to uncorrelated sources are orthogonal, so we will use the whitening step in these simulations for consistency. In the simulations, whitening is carried out <u>after</u> the addition of noise as would happen in practice.

Note that normalisation of the Gram-Schmidt components is carried out when whitening and these components are shown in Figure 13 below.

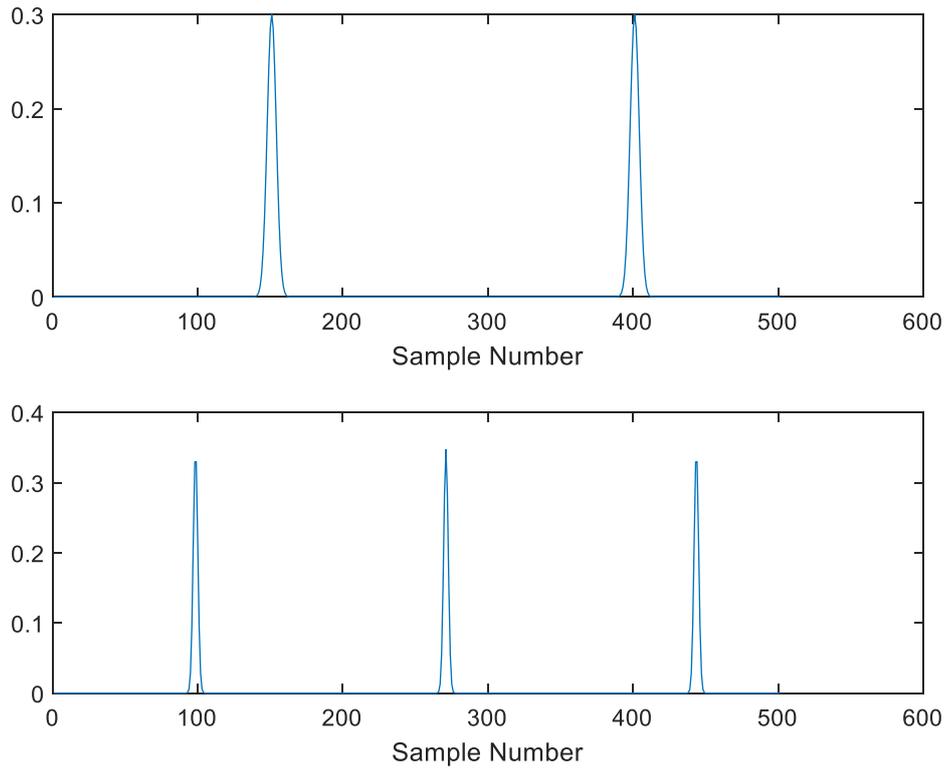

*Figure 13 – Whitened Components corresponding to the Sources shown in Figure 1*

Because of the normalisation, the whitened component corresponding to Source 2 (lower plot) now has a higher peak than the normalised component corresponding to Source 1 (upper plot). Hence when used in the Maximum method it is Source 2 that will tend to be detected first.

First, let us look at the case where the standard deviation of the noise level is 1% of the peak value of the original Source 2 (Figure 1) i.e. noise sd = 0.1 x 0.01 = 0.001.

In the following figures, the rms estimation errors as a function of sample number are shown in cyan for the PCA and red for the Maximum method. As a reference the clean normalised source is shown in purple in the same figure.

At the top of Figure 14 it can be seen that, for Source 1, the rms estimation errors for the PCA and Maximum method are negligible compared with the actual sources shown in purple. The corresponding errors for Source 2 (bottom of Figure 14) are more significant when compared with the actual source, which is understandable as this is the smaller amplitude source (see Figure 1).

**RMS Errors Source 1**

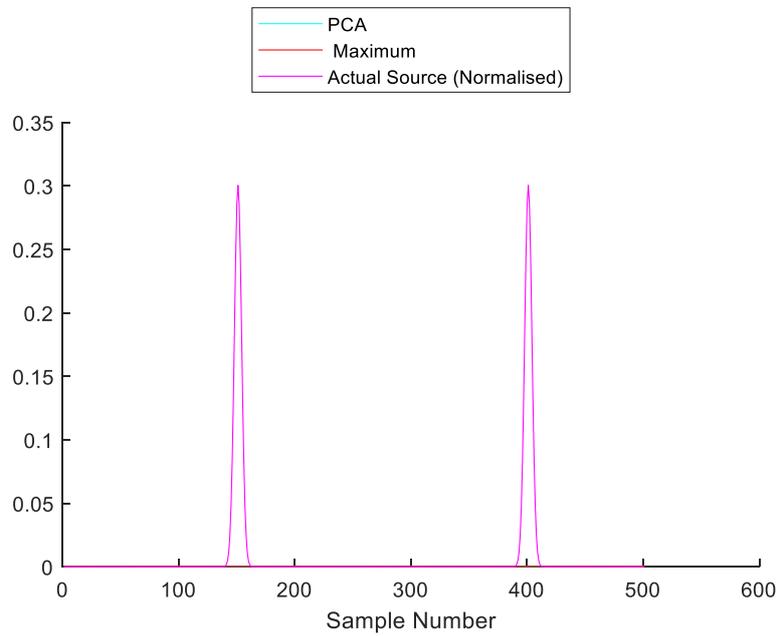

**RMS Errors Source 2**

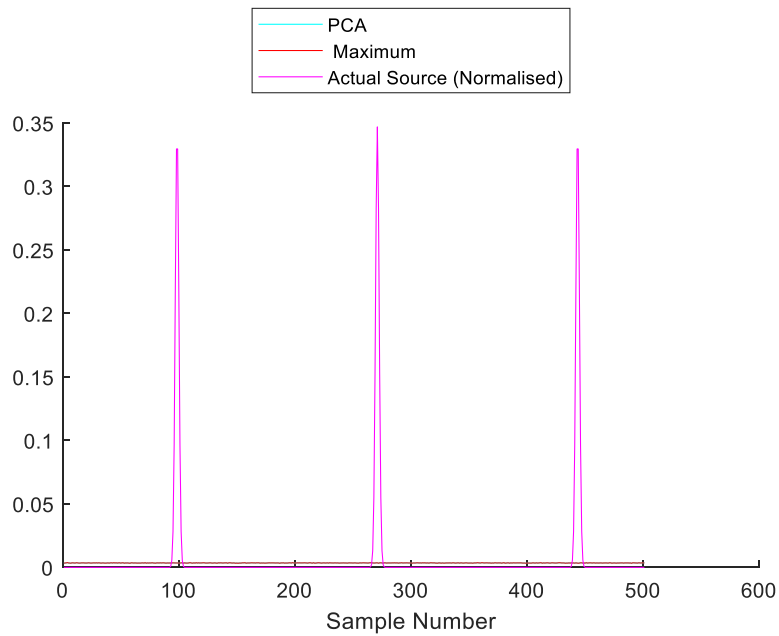

*Figure 14 – Comparison of RMS estimation errors between PCA and the Maximum methods. Cyan: PCA, Red: Maximum method and Purple: actual source. Top graph: Source 1 errors, Bottom graph: Source 2 Errors.*

There is no observed significant difference in performances of the PCA and Maximum methods for this noise level.

In Figure 15, we show an alternative way to illustrate the working of the Maximum method when there is noise present. A phase plot is shown where the direction corresponding to the first estimated source (blue) is superposed for all Monte Carlo runs on top of the phase plot of the clean whitened signal (red). Note that because, after normalisation, Source 2 has a larger peak value than Source 1, then it is Source 2 that is preferentially detected by the Maximum method. There is very little deviation of this direction over all Monte Carlo runs from the vector corresponding to Source 2.

**Phase Plot for the Maximum Method**

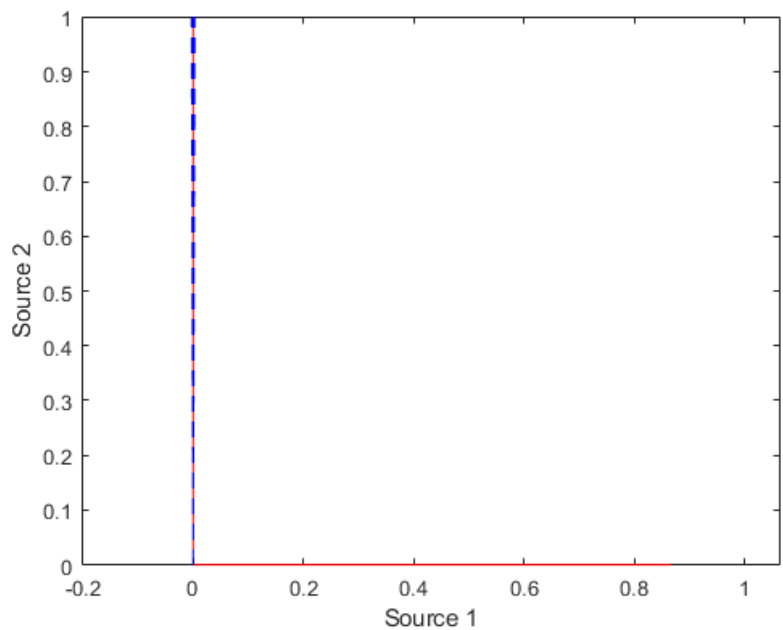

*Figure 15 – Phase plot of first estimated source (blue) over all Monte Carlo runs superposed on top of the phase plots of the clean signal (red)*

We now look at the case where the standard deviation of noise is increased to 5% of the amplitude of the smaller sources, that is 0.05*0.1 = 0.005.

It can be seen in Figure 16, that the rms errors for the larger amplitude Source 1 are still negligible, but the rms errors for Source 2 now show peak values at the points where Source 2 is a maximum. This occurs in equal measure for both the PCA and Maximum methods.

**RMS Errors Source 1**

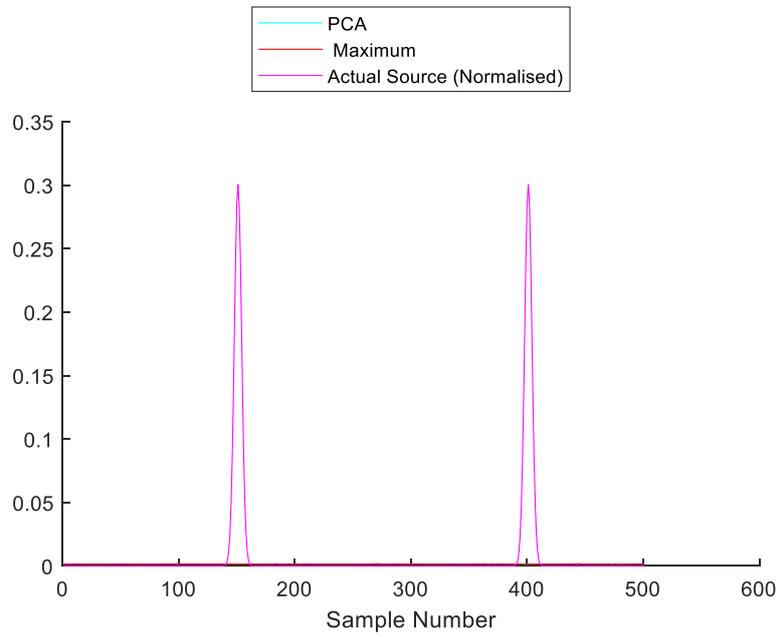

**RMS Errors Source 2**

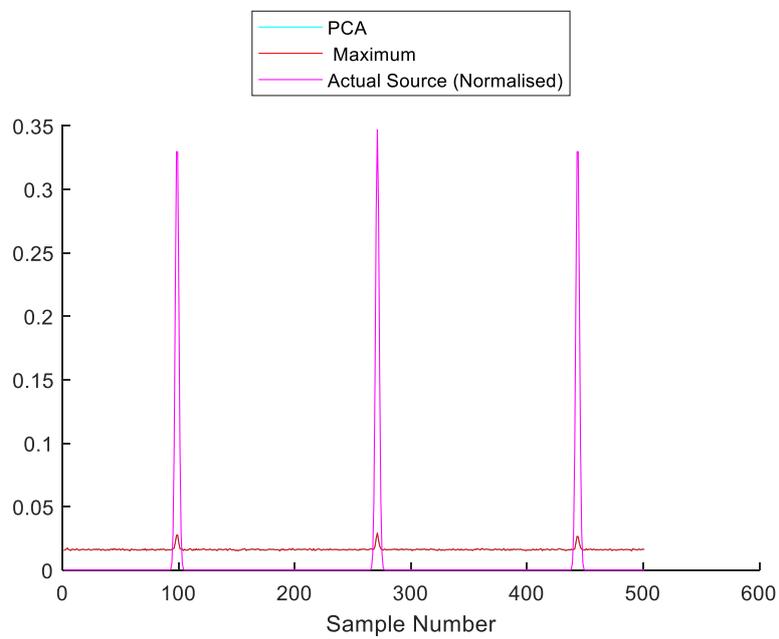

*Figure 16 – Comparison of RMS estimation errors between PCA and the Maximum methods. Cyan: PCA, Red: Maximum method and Purple: actual source. Top graph: Source 1 errors, Bottom graph: Source 2 Errors.*

If one now looks at the phase plot for the Maximum method, see Figure 17, we can see that the principal direction estimated for Source 2 differs significantly from the actual direction; there is thus a small but noticeable component of the principal direction in the direction of Source 1.

**Phase Plot for Maximum Method**

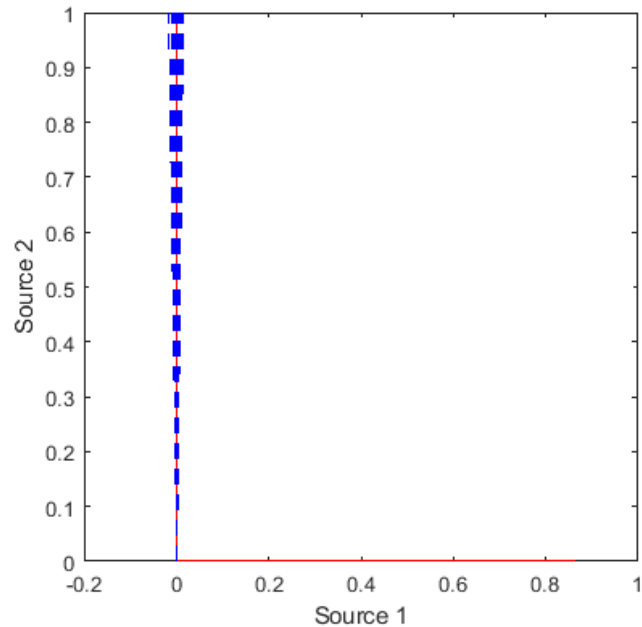

*Figure 17 – Phase plot of first estimated source (blue) over all Monte Carlo runs superposed on top of the phase plots of the clean signal (red)*

Now let us increase the standard deviation of noise to be 7.5% of the amplitude of Source 2 = 0.075 x 0.1 = 0.0075.

**RMS Errors Source 1**

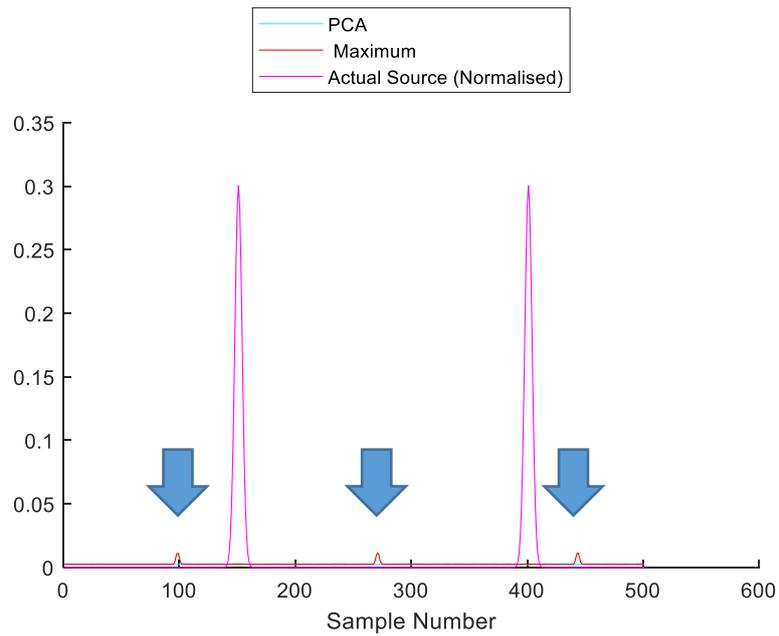

**RMS Errors Source 2**

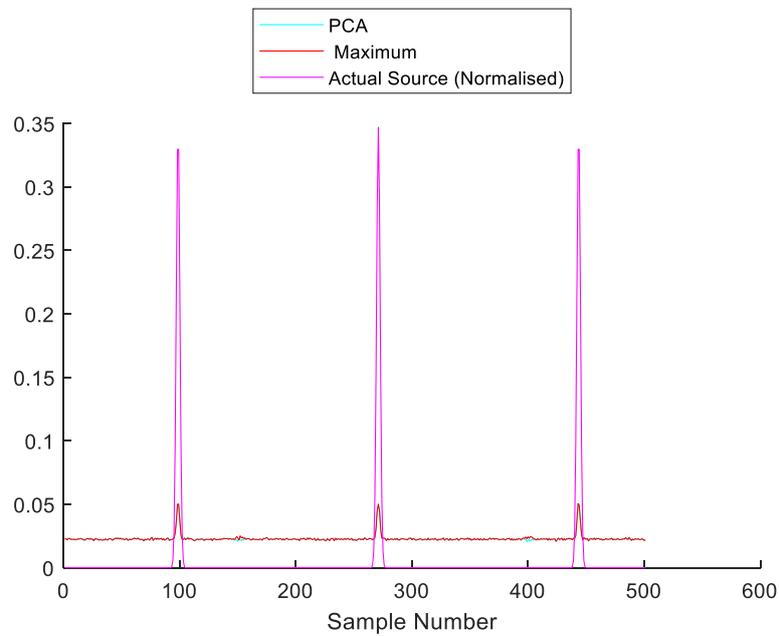

*Figure 18 – Comparison of RMS estimation errors between PCA and the Maximum methods. Cyan: PCA, Red: Maximum method and Purple: actual source. Top graph: Source 1 errors, Bottom graph: Source 2 Errors.*

The estimates of Source 1 are contaminated by contributions to Source 2 as shown by the arrows in the top half of Figure 18; this can be explained by looking at the phase plot in Figure 19. As noise increases, the estimated direction in the phase plot corresponding to Source 2 varies more than before, but now, in some cases, Source 1 is detected first. Because noise affects normalised Source 2 more than normalised Source 1 there is more fluctuation of the principal direction about the x-axis corresponding to Source 1. This explains why the estimates of Source 1 using the Maximum method pick up significant contributions from Source 2 as indicated by the arrows in the top half of Figure 18; this problem is not observed when using PCA.

**Phase Plot for Maximum Method**

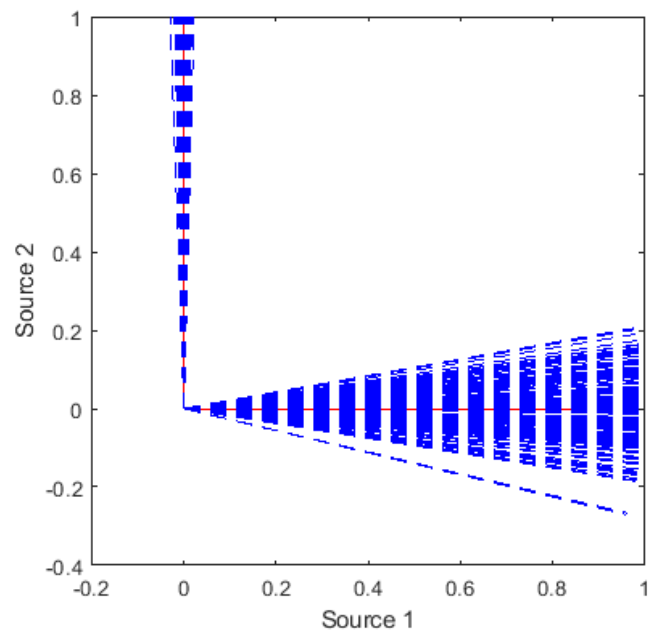

*Figure 19 – Phase plot of first estimated source (blue) over all Monte Carlo runs superposed on top of the phase plots of the clean signal (red)*

Finally, let us look at the case where the noise standard deviation Noise sd is 10% of amplitude of Source 2 = 0.1*0.1 = 0.01.

It can be seen in Figure 20 below that the errors in the estimates for Sources 1 and 2 are contaminated by the other source as indicated by the arrows.

**RMS Errors Source 1**

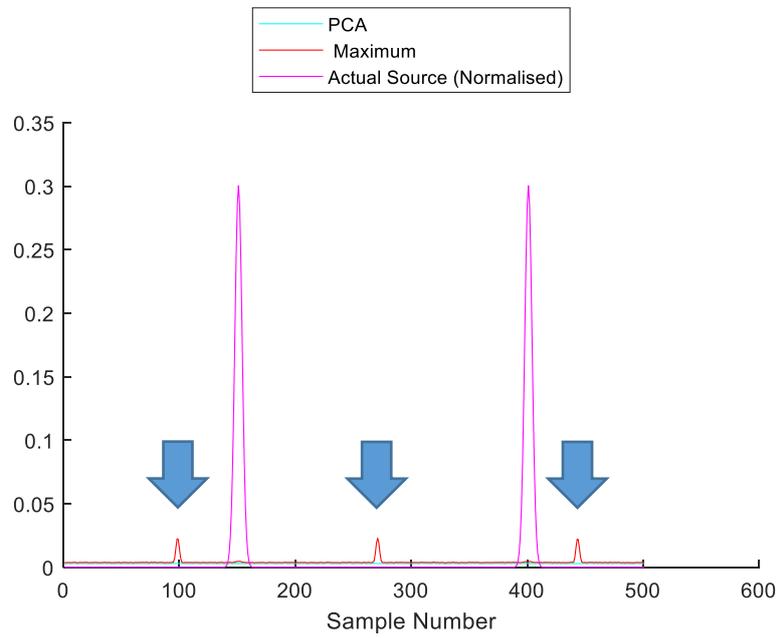

**RMS Errors Source 2**

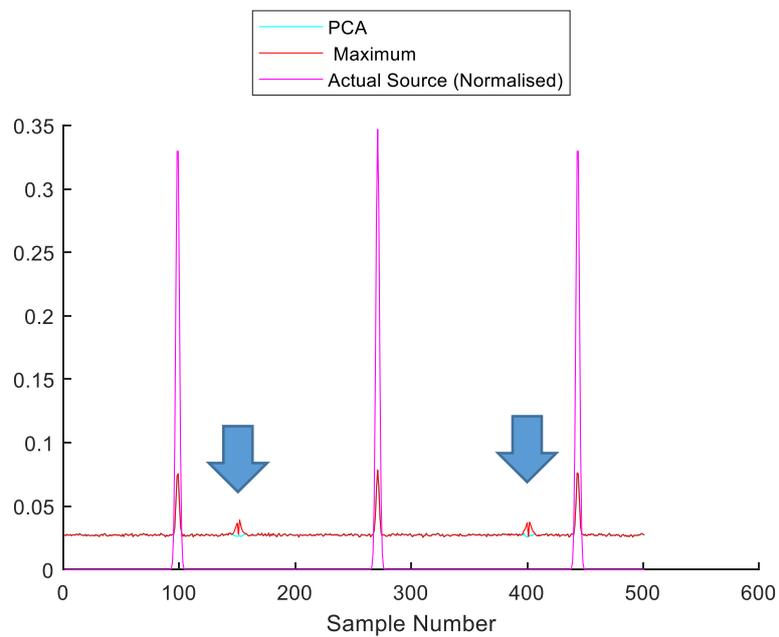

*Figure 20 – Comparison of RMS estimation errors between PCA and the Maximum methods. Cyan: PCA, Red: maximum method and Purple: actual source. Top graph: Source 1 errors, Bottom graph: Source 2 Errors.*

The phase plot showing the direction of the first component and the actual component directions is illustrated in Figure 21 below:

**Phase Plot for Maximum Method**

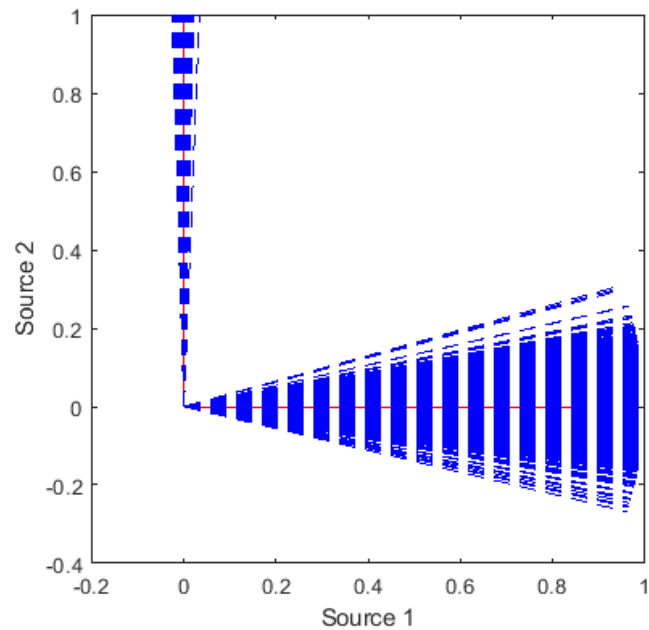

*Figure 21 – Phase plot of first estimated source (blue) over all Monte Carlo runs superposed on top of the phase plots of the clean signal (red)*

As for the previous noise level, when using the Maximum method, Sources 1 and 2 can both be detected first and the estimates of each source is affected by the other source as indicated by the arrows in Figure 20.

In this situation, for the estimate of Source 1, PCA is more stable to increasing noise than the Maximum method; the contamination of Source 2 to the estimates of Source 1 indicated by the arrows in the top figure is not observed for PCA (cyan) but is noticeable for the Maximum method (red).

To summarise, there are comparable performances between PCA and Maximum methods for noise up to 5% of the amplitude of the smaller source; PCA gives better estimates for the larger source for noise levels above this value.

It should be noted that PCA is a statistical technique that processes the whole data through the covariance matrix. The Maximum method relies on using the maximum value ignoring the other data, which can make it more sensitive to noise.

## 4.2 Mixture of Two Correlated Sparse Sources

In Section 4.1, we looked at the case of a mixture of two uncorrelated sources and showed that the Maximum method is able to estimate each source to within a scaling constant.

In this Section, we compare the application of the PCA and Maximum methods to a mixture of two correlated sparse sources. The original sources are shown in Figure 22 and the mixture in Figure 23 is obtained using the mixing matrix:

$$A = \begin{pmatrix} 6.5 & 1 \\ 3 & 1 \end{pmatrix}$$

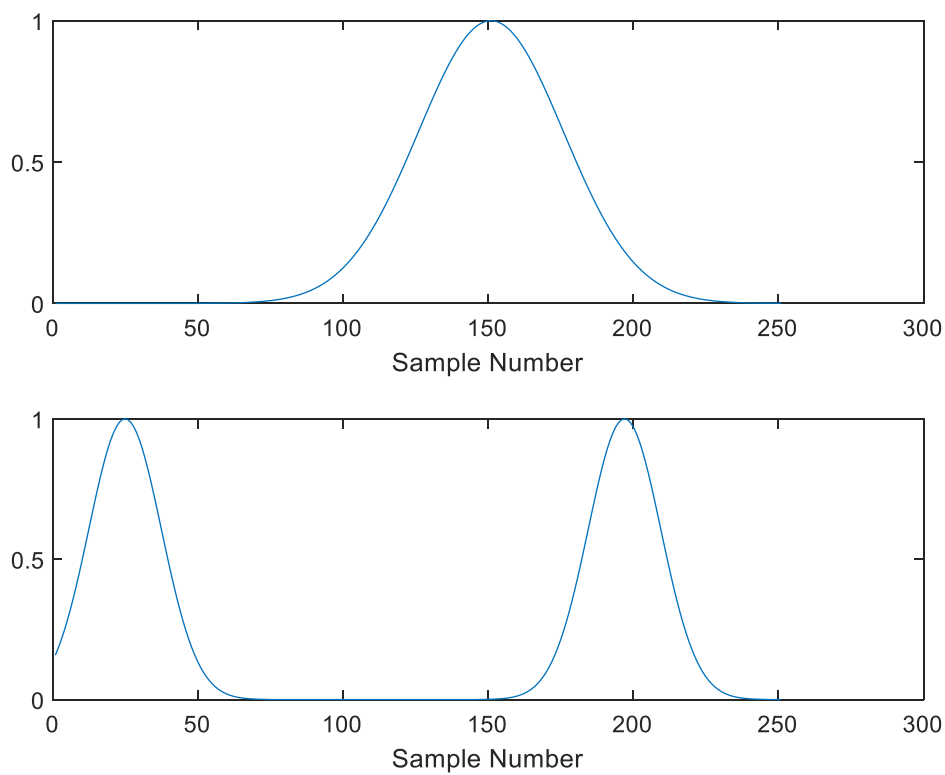

*Figure 22 –Two correlated sparse sources*

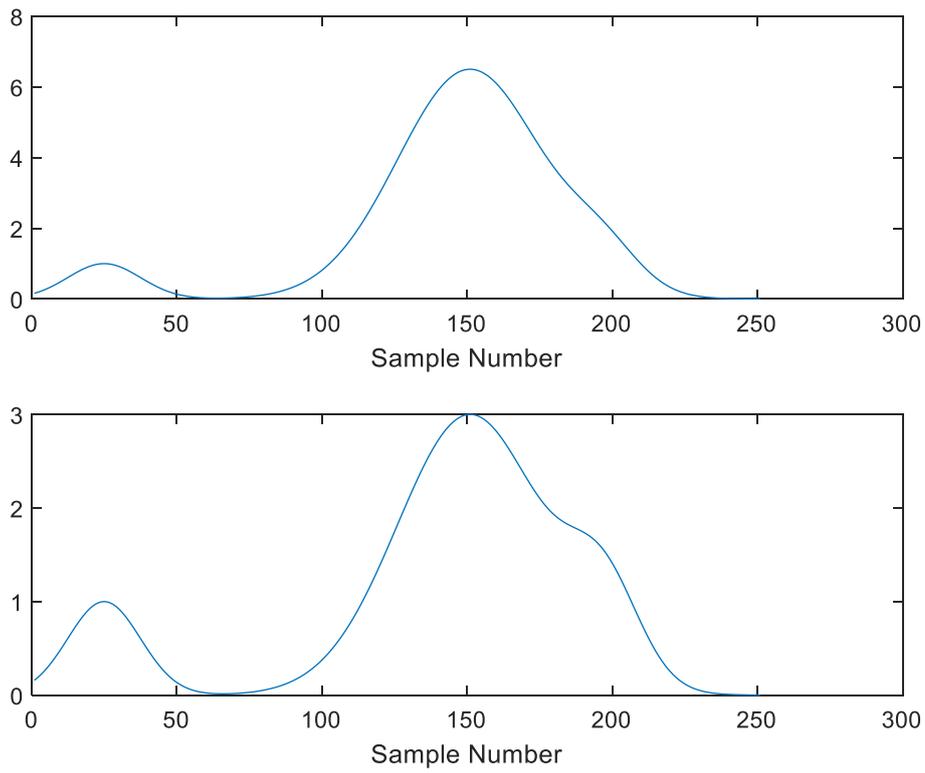

*Figure 23 –Mixture of two correlated sparse sources*

The estimated sources using the Maximum and PCA methods are shown in Figures 24 and 25 respectively.

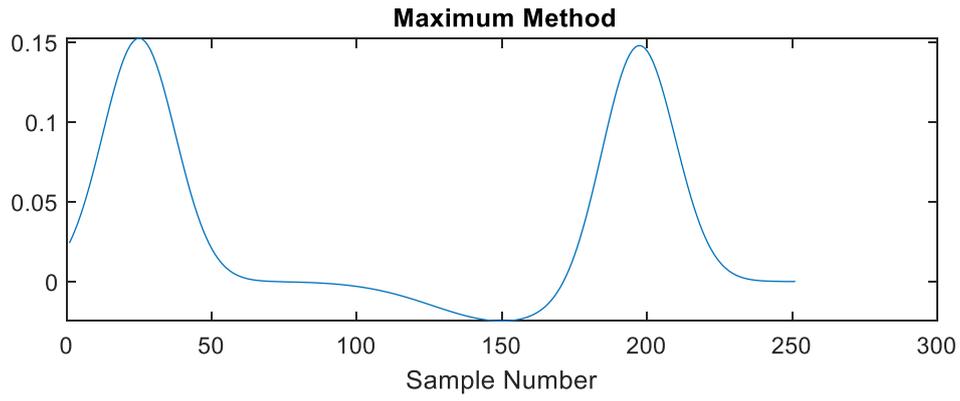
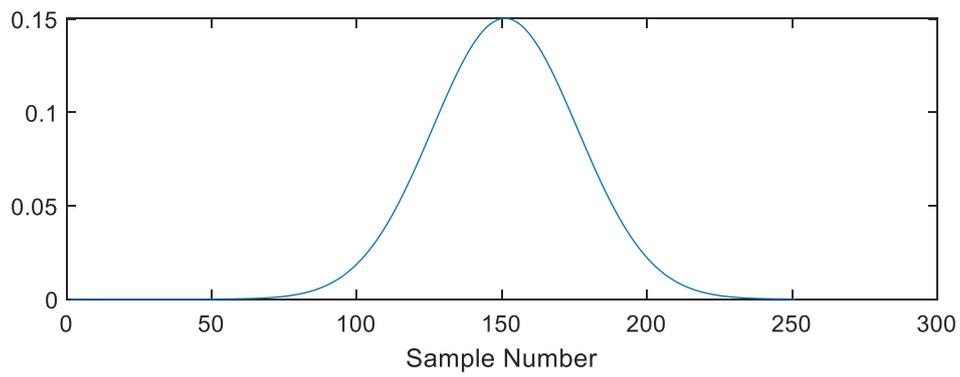

*Figure 24 – Sources estimated using the Maximum Method*

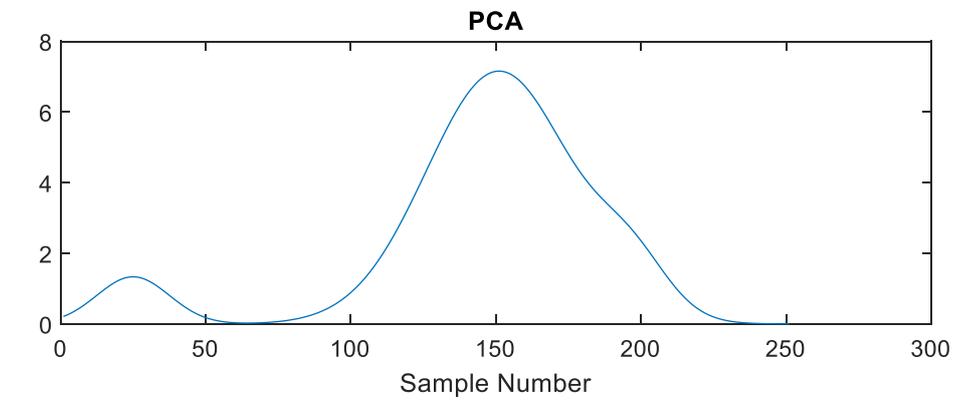
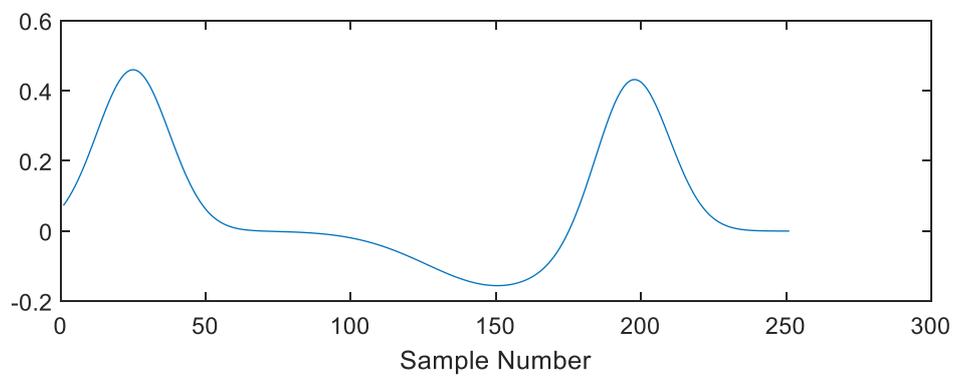

*Figure 25 – Sources estimated using PCA*

In Figure 24, it can be seen that, when using the Maximum method, the first estimated source is contaminated by the other source.

In the phase plot shown in Figure 26, the Gram-Schmidt normalised components are shown in red and the principal direction found using the Maximum method is plotted in blue.

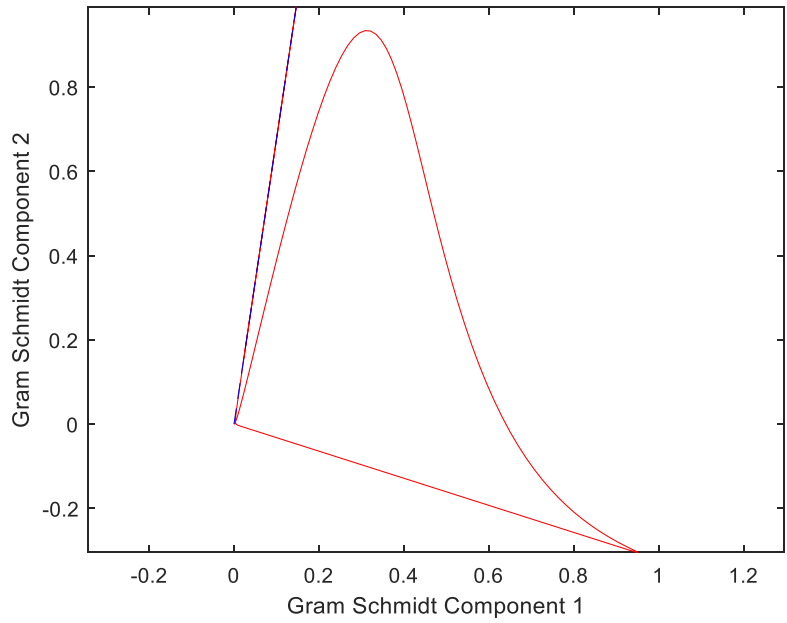

*Figure 26 – Phase Plot of Normalised Gram-Schmidt Components (red) with direction of first estimated source detected by the Maximum method shown in blue*

It can be seen that the Maximum method correctly detects the direction of one of the sources. However, because the sources are now correlated, the directions corresponding to the two sources are not orthogonal. Hence, the first estimated sources will be contaminated by the other source as observed in the top graph in Figure 24.

However, after applying deflation, the second source is estimated to within a scaling constant, in agreement with Equation (23).

When PCA is applied, the estimated sources are as shown in Figure 25, where it can be seen that each estimated source is contaminated by a contribution from the other source. The eigenvectors are plotted as red and yellow lines in Figure 27 below:

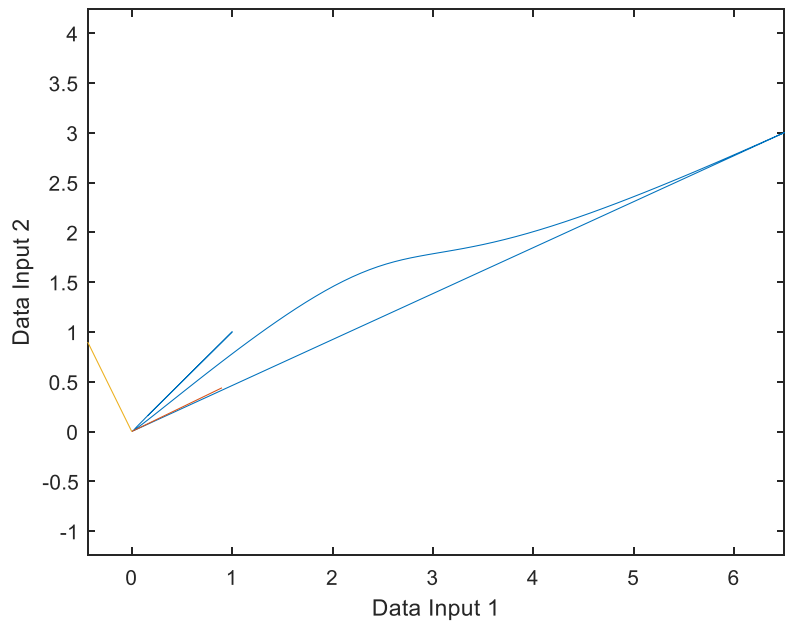

*Figure 27 – Phase plot of data mixtures (Blue) with direction of eigenvectors shown in red and yellow*

One of the directions (red) corresponds to one of the sources; however, there is a component of the second source in this direction. The second direction (yellow) corresponding to the eigenvector with the lower eigenvalue picks up contributions from both sources.

The advantage of the Maximum method over the PCA method for correlated sparse sources is that the former locally looks at the phase plot and detects the point corresponding to a source. The PCA determines the direction of maximum variance, processing all the data points, and this direction does not necessarily correspond to one of the sources.

## 4.3    Mixture of Two Correlated Sources with Coincidence of Peaks

In this section, we compare the Maximum and PCA methods to a mixture of two sources where there is a coincidence of peaks from the two sources as indicated by arrows in Figure 28.

**Original Sources**

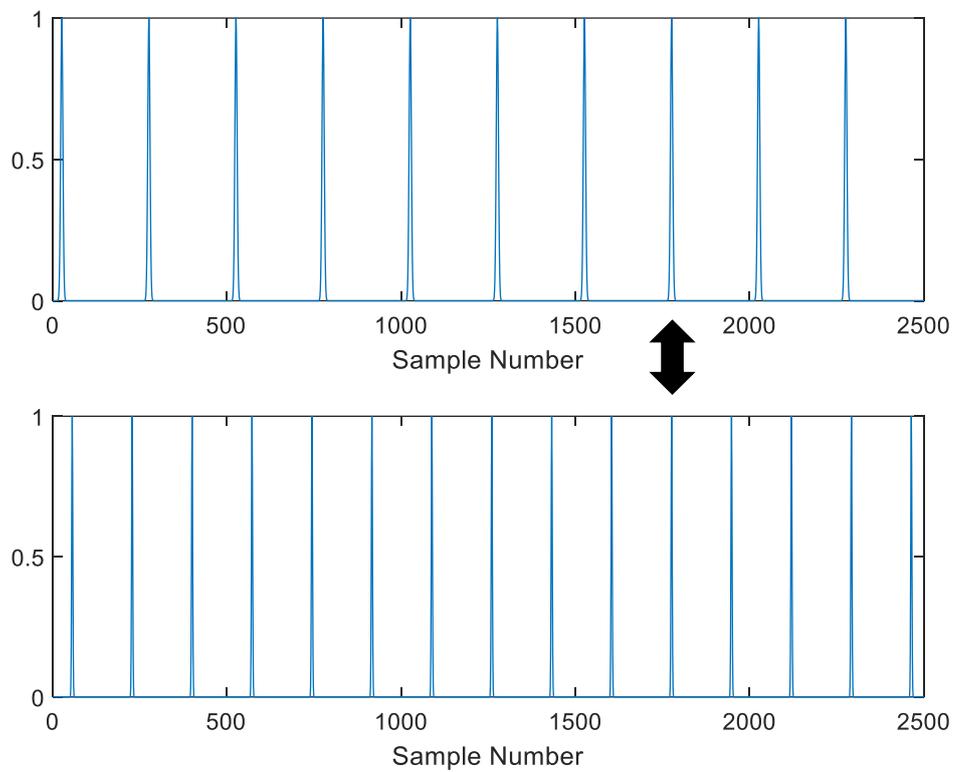

*Figure 28 – Two sparse sources with coincidence of peaks*

The mixed signal, using the mixing matrix

$$A = \begin{pmatrix} 6.5 & 1 \\ 3 & 1 \end{pmatrix}$$

is shown in Figure 29.

**Mixture**

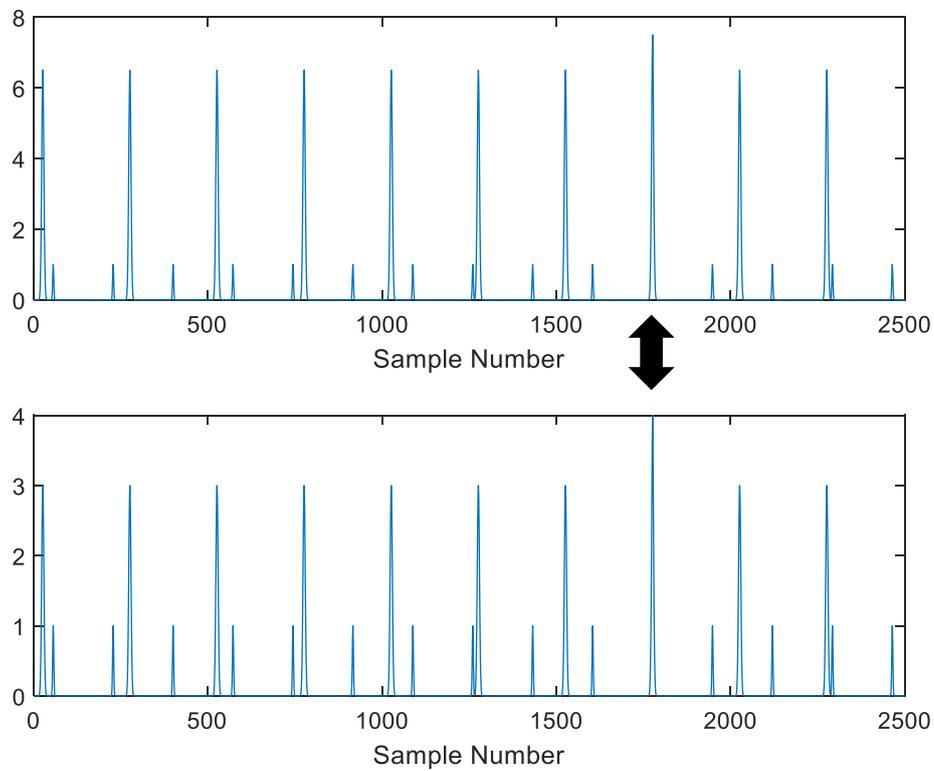

*Figure 29 – Mixture of Sparse Sources in Figure 28*

The estimates of the two sources when applying the Maximum method is shown in Figure 30 with the corresponding result for the PCA shown in Figure 31. It can be seen that the PCA provides by far the better estimates, with the Maximum method breaking down

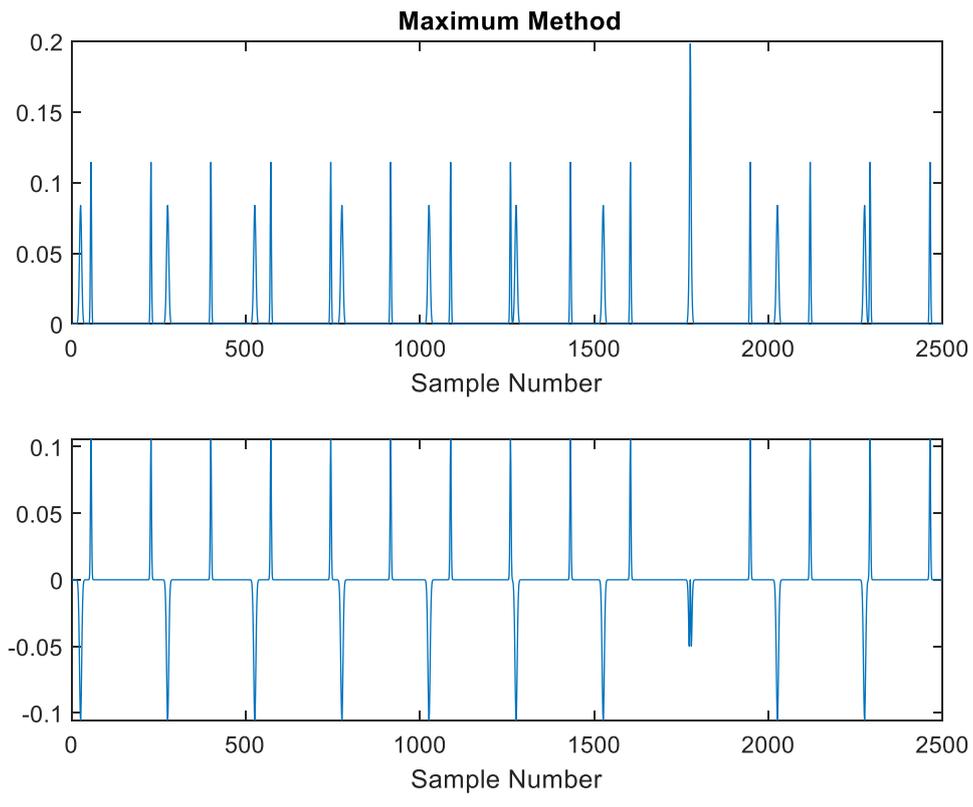

*Figure 30 – Sources estimated using the Maximum method*

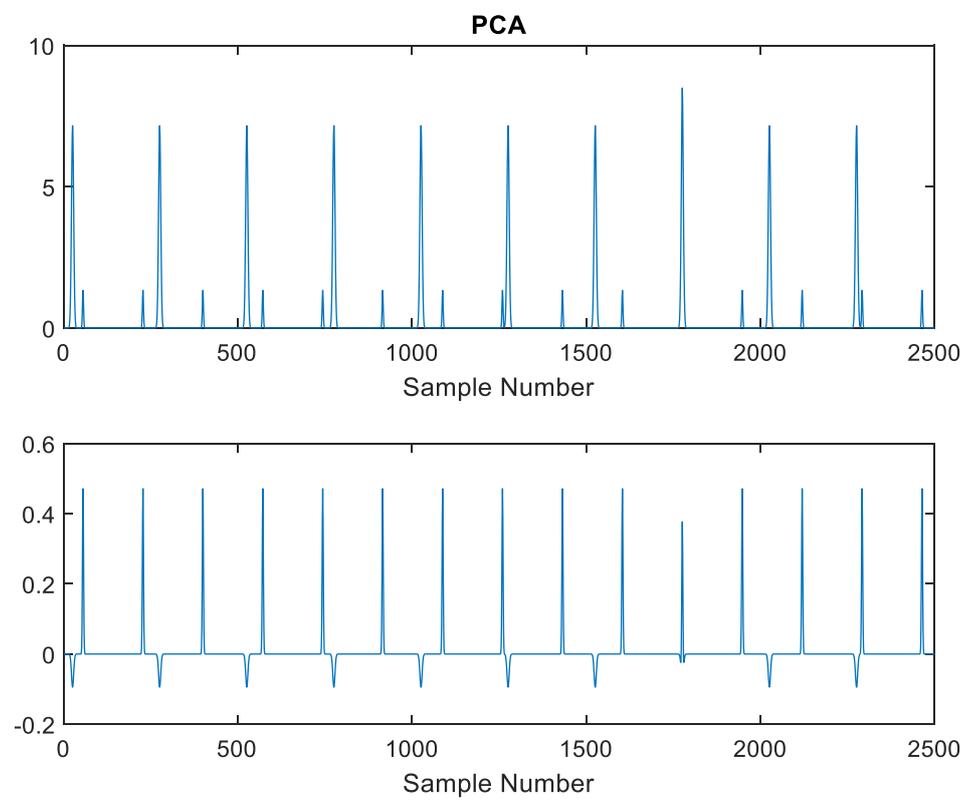

*Figure 31 – Sources estimated using PCA*

The reason for the poor performance of the latter method can be seen in Figure 32, where the Gram-Schmidt whitened components are plotted against each other.

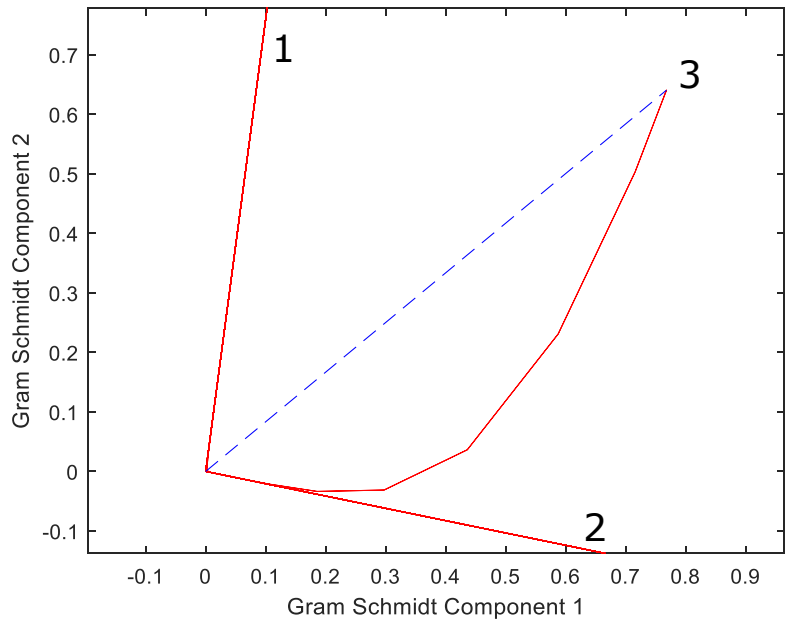

*Figure 32 – Phase plot of normalised Gram-Schmidt components (red) and principal direction detected using the Maximum Method (blue)*

The sources are represented in the phase plot by the lines 1 and 2. The curved line 3 comes from the co-incidence of the peaks indicated by the arrows in Figure 29. As the two sources reinforce at this time plot this contributes to a maximum in the phase plot indicated by the dashed blue line in Figure 32, which does not correspond to one of the sources; it is this direction that is detected, erroneously, by the Maximum method.

PCA has a better performance because it processes the whole data and is relatively insensitive to coincidences of peaks.

Phase space methods can be applied to this problem, but there needs to be a method to detect the straight lines in the phase plot (e.g. 1 and 2 in Figure 32) and not just look for a maximum; methods that can do this are described in [25].

In general, as the Maximum method relies on finding a single point in phase space, it is more likely to be sensitive to outliers and coincidence of peaks than the PCA method.

## 4.4 Estimation of Fetal and Maternal Components of Abdominal ECG Data

One popular application of BSS and PCA methods is to the extraction of fetal and maternal ECG signals from abdominal ECG data. A recent comparison of the two approaches is described in [28].

In this section, we compare the PCA and Maximum methods to the analysis of selected datasets to test the feasibility of using the simpler Maximum method.

We will be using three sets of data. Two sets of data are taken from the Daisy Database [29]. This data consists of ECG signals taken from an expectant mother. The data consists of 8 leads, 1 to 5 being abdominal and 6 to 8 thoracic. The first 1000 samples of the data are chosen.

Data from a typical abdominal lead is shown in Figure 33.

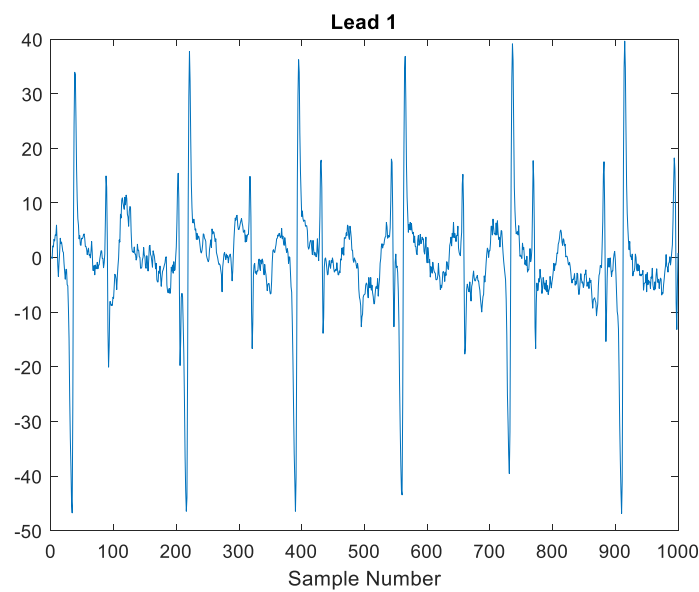

*Figure 33 – Lead 1 Abdominal Signal taken from [29]*

The second set of data is a subset of this database, just leads 1 to 4, which are abdominal only and hence each consists of a mixture of maternal and fetal ECG signals.

The third set of data is taken from the Physionet database [30], Abdominal and Direct Fetal ECG Database [31,32]. The first 3000 samples of file r01.edf in this database are analysed.

When applying the PCA, the data are not normalised to avoid the problem of fixing the values of the eigenvectors as mentioned in Section (2).

When applying the PCA and BSS methods to experimental data of course we cannot say for certain how accurate each estimated source is as we do not know exactly what the actual maternal and fetal components are. Visual comparison can be subjective. In the end, what we are looking for is an estimate of the fetal and maternal components which have clear R-waves above background noise that can be used for (i) estimation of the instantaneous heart rate and (ii) reference points that can be used to obtain averaged maternal and fetal ECG signals.

### 4.4.1    8-lead Data [29]

The result of applying PCA to the data is shown in Figure 34.

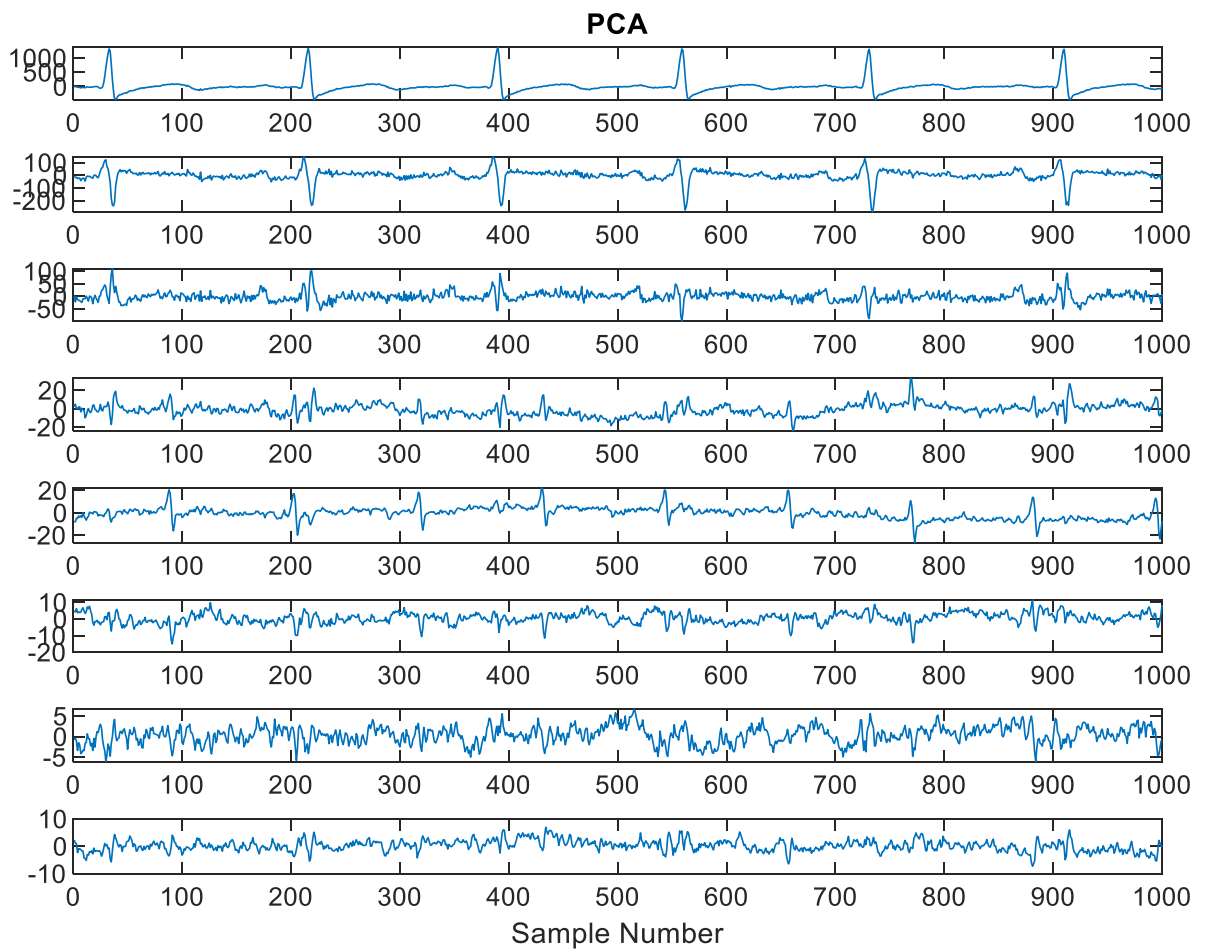

*Figure 34 – Sources Estimated using PCA*

Clear maternal contributions can be seen in the first and second components, with the fetal contribution prominent in the 5$^{th}$ component.

If we apply the Maximum method to the raw data, without performing pre-whitening, the resulting estimated sources are shown in Figure 35:

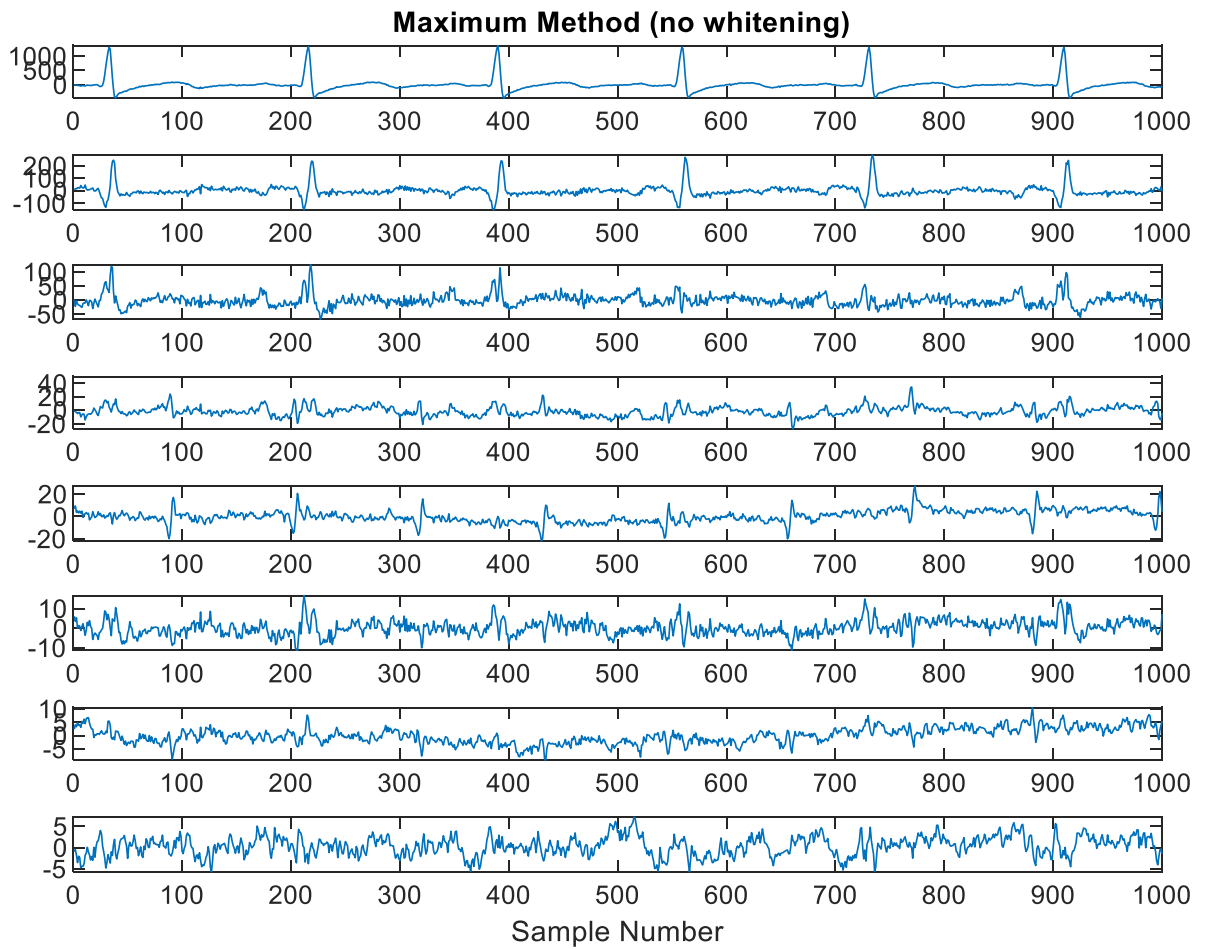

*Figure 35 – Sources Estimated sing the Maximum method (no pre-whitening)*

Comparing Figures 34 and 35, it can be seen that the Maximum method produces very similar estimated sources compared with the PCA.  This is to be expected, as the direction of maximum variance, determined by PCA, will be close to the direction of the maximum of the trajectory in the phase space.

If we calculate the correlation coefficients between corresponding sources obtained by the two methods, then the results are as shown in Table 1 below.  Note that for the last two sources, the highest correlations are obtained between sources 7 for the PCA and 8 for the Maximum method and 8 for the PCA and 7 for the Maximum method.  The high correlation between corresponding sources, particularly the first five sources found by each method, illustrates the similarity between the Maximum method (without whitening) and PCA for this dataset.

| PCA Source Number | Maximum Source Number | Correlation (4 d.p.) |
|---|---|---|
| 1 | 1 | 1 |
| 2 | 2 | -0.9964 |
| 3 | 3 | 0.8978 |
| 4 | 4 | 0.8401 |
| 5 | 5 | -0.9210 |
| 6 | 6 | 0.6781 |
| 7 | 8 | 0.9230 |
| 8 | 7 | -0.5766 |

*Table 1 – Correlation Coefficients Between Sources Estimated using PCA and the Maximum method*

It should be noted that the final two estimated sources appear to consist mainly of noise.

If we apply the Maximum method, but this time applying pre-whitening using the Gram-Schmidt method, then the estimated sources are as shown in Figure 36:

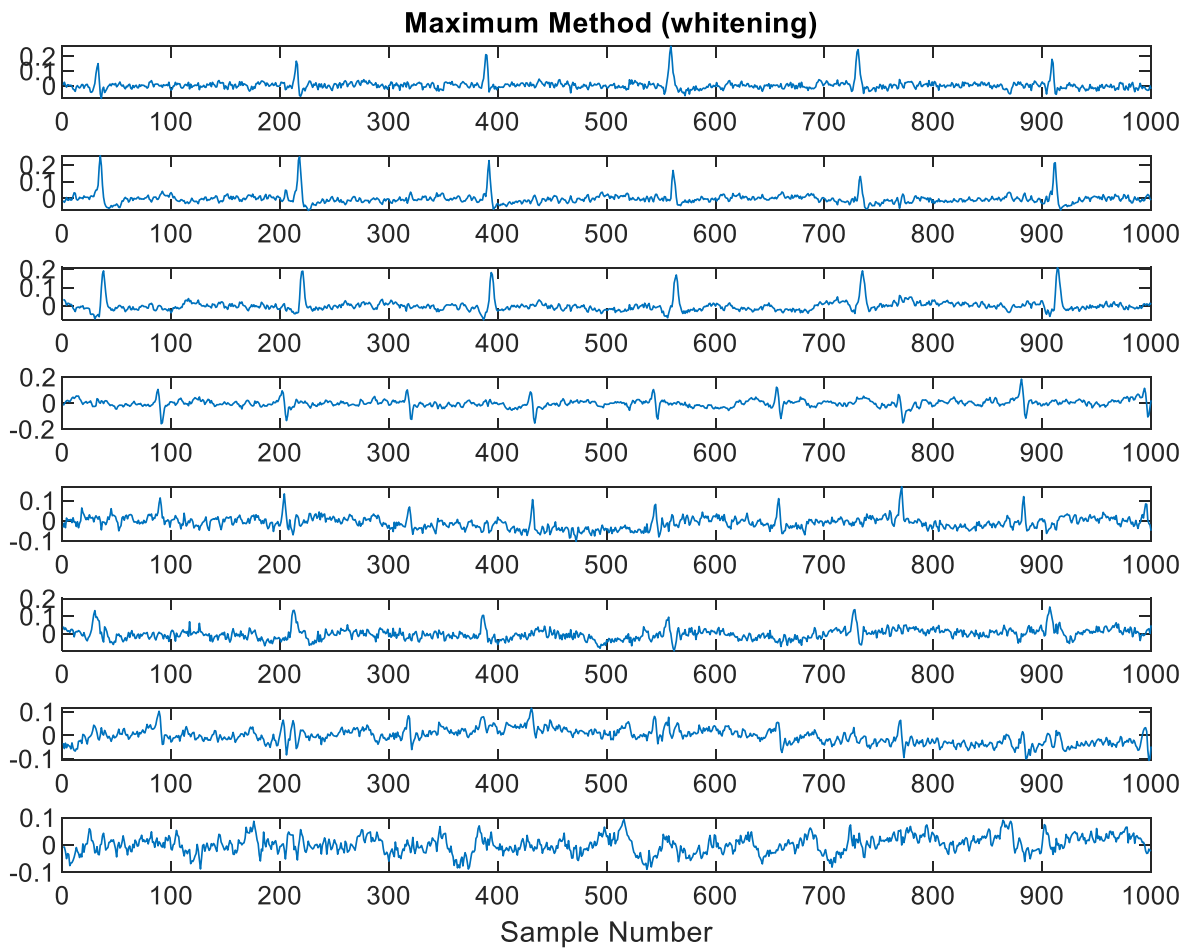

*Figure 36 – Sources Estimated sing the Maximum method with whitening*

There are now two fetal components detected, estimates 4 and 5, with the fourth estimated source of better quality. However, the maternal components, 1,2 and 3 are of poorer quality (more noise) than when no whitening is carried out; this is not a problem for this particular dataset, as leads 6 – 8 are thoracic leads and so will contain purely maternal signals. Hence, the main aim of the analysis of this particular dataset is to extract the fetal contribution to the abdominal signals.

Now whitening using Gram-Schmidt, in theory, should yield better maternal and fetal components as the principal directions in phase space corresponding to these contributions should be as close to 90 degrees as possible.

It is of interest to compare our results with those of the FastICA and these are shown in Figure 37 below where pow3 non-linearity is used and deflation is employed to estimate each source iteratively.

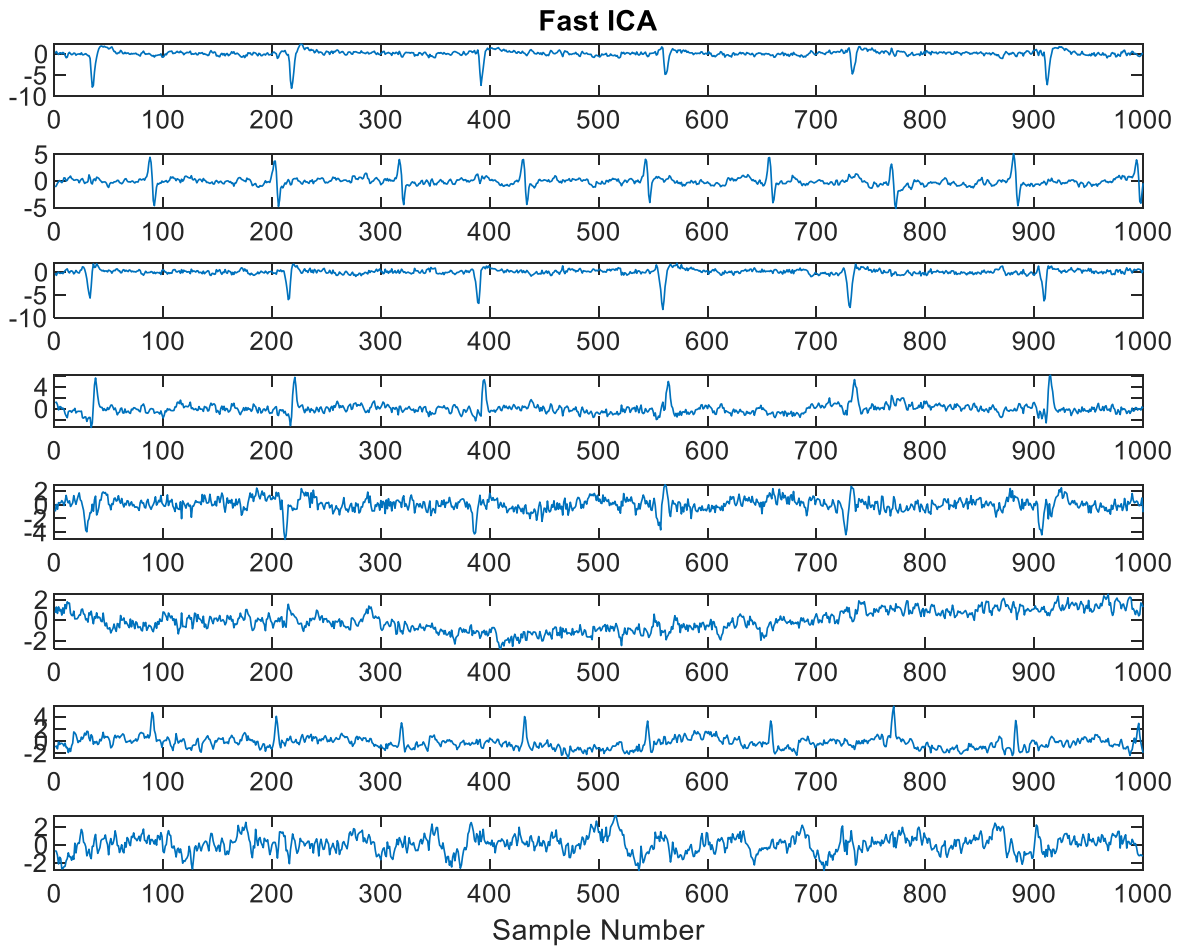

*Figure 37 – Sources Estimated using FastICA*

If we compare this result with Figure 36 then we can see that there is no noticeable improvement in maternal estimation compared to the Maximum method. There are two estimated fetal components (estimates 2 and 7) when using the Fast ICA, like with the Maximum method with pre-whitening. It should be noted that the pre-whitening step used when applying the FastICA is to apply PCA first and then normalise the PCA estimates, rather than using the Gram-Schmidt method.

Both the Maximum method and PCA give comparable results to those obtained using the FastICA when estimating the fetal contribution to the abdominal signals. It should be noted that the above observations apply to this particular data set; we will later look at other datasets to see if similar results are observed.

Finally, let us look more closely at the fetal components estimated using the PCA, Maximum Method (no whitening), Maximum method (whitening) and FastICA. These are as shown in Figure 38 below

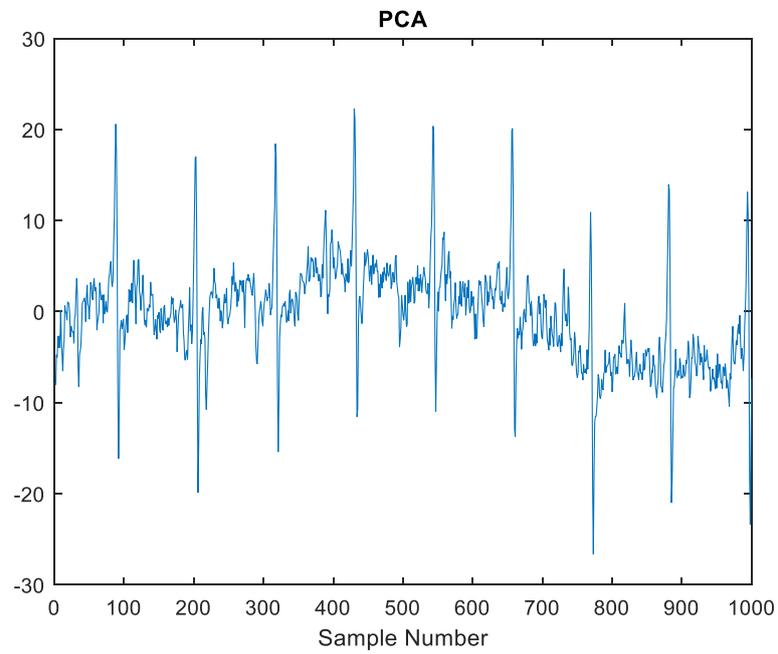

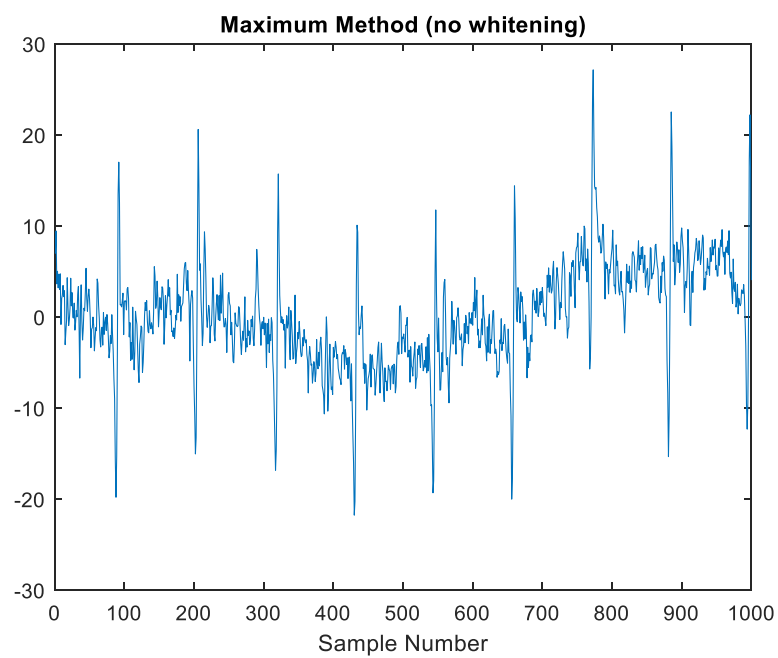

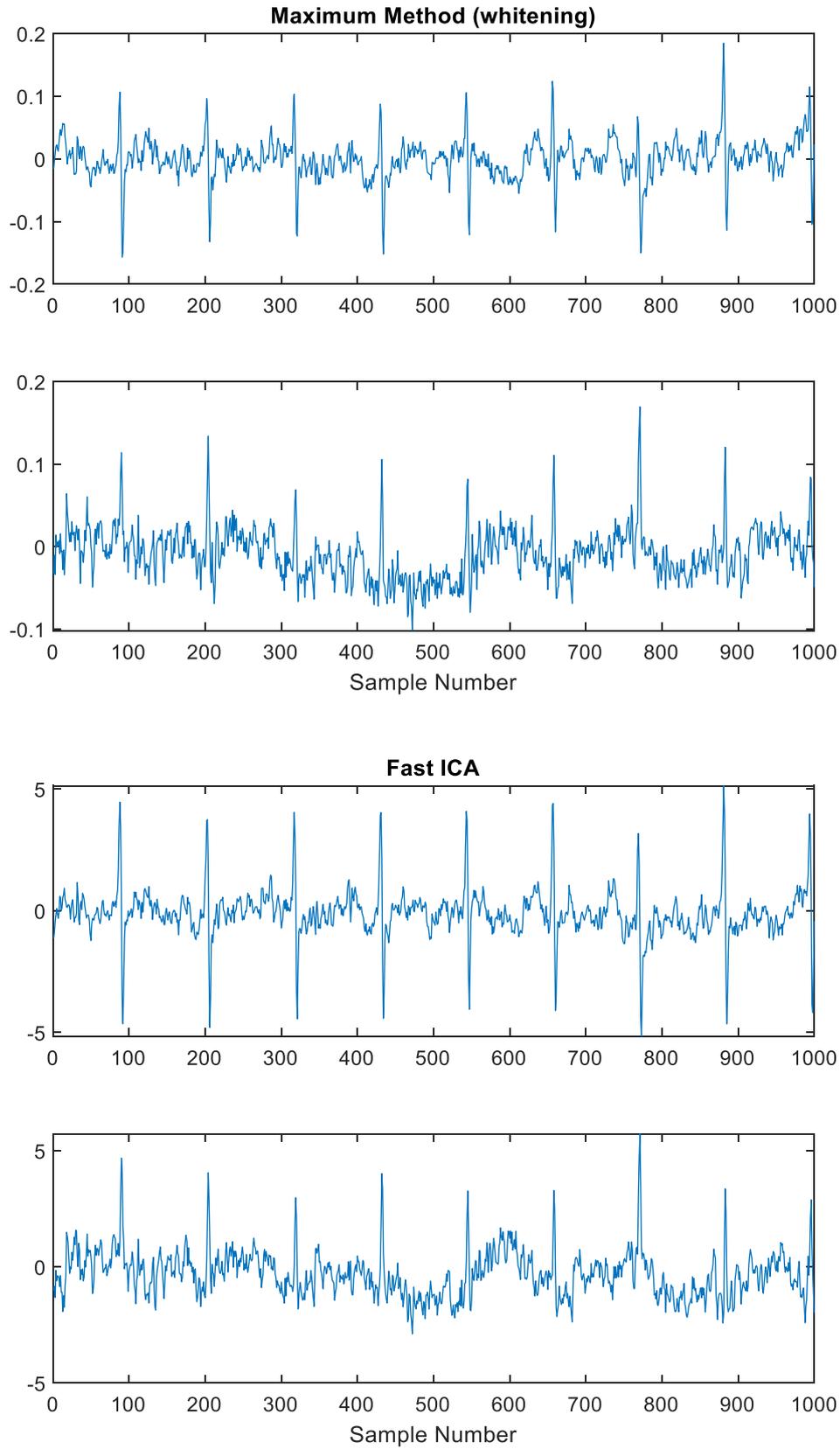

*Figure 38 – Comparison of PCA, Maximum-no whitening, Maximum with whitening and Fast ICA for 8-lead data*

PCA and the Maximum method (no pre-whitening) give very similar fetal estimates, which is to be expected given the close relation between the two methods.

Two dominant fetal source estimates are obtained with both the Maximum method with pre-whitening and Fast ICA. For each method, the first estimated fetal source is of higher quality than the second estimated fetal source. The best fetal estimate looks less noisy for FastICA than the Maximum method with pre-whitening; however, both estimates could be used for determining the fetal R-wave positions.

### 4.4.2 4-Lead Data

We now look just at leads 1 to 4 of the data in the Daisy database [29] and take just the first 300 samples; these data are shown in Figure 39 below:

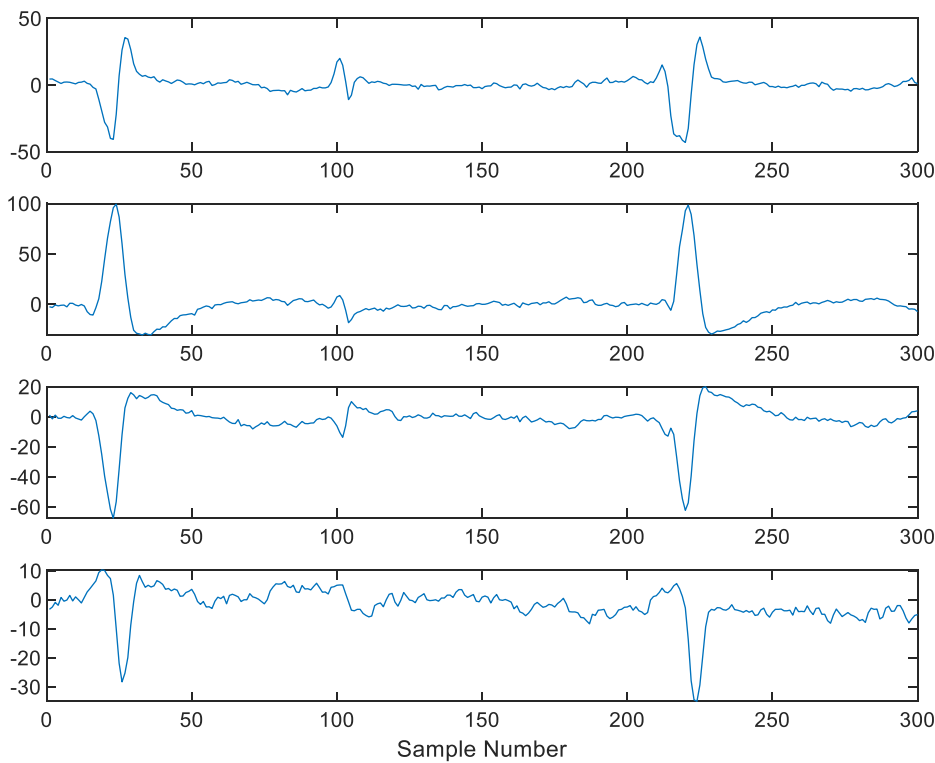

*Figure 39 – Leads 1 to 4 of Abdominal Data [29]*

This is a more challenging problem than the 8-lead data in section 4.4.1 as each set of data contains a mixture of both maternal and fetal ECG signals. The result of applying PCA to this set of data is shown in Figure 40 below:

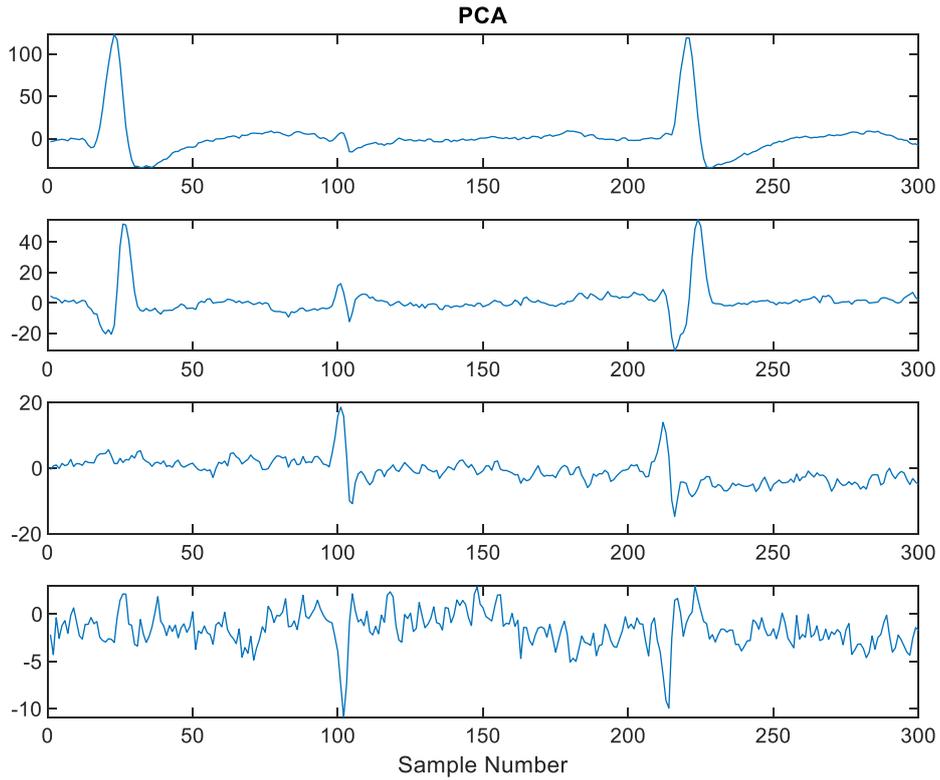

*Figure 40 – Estimated Sources using PCA*

The principal component (top graph) contains mainly maternal with a small fetal contribution. The third component contains a clear fetal component with no significant maternal contribution – note that the second fetal peak is coincident with the second maternal peak in the original data, but this does not pose a problem for the PCA method.

If we now apply the Maximum method with whitening (using the Gram-Schmidt method), the estimated sources are as shown in Figure 41 below:

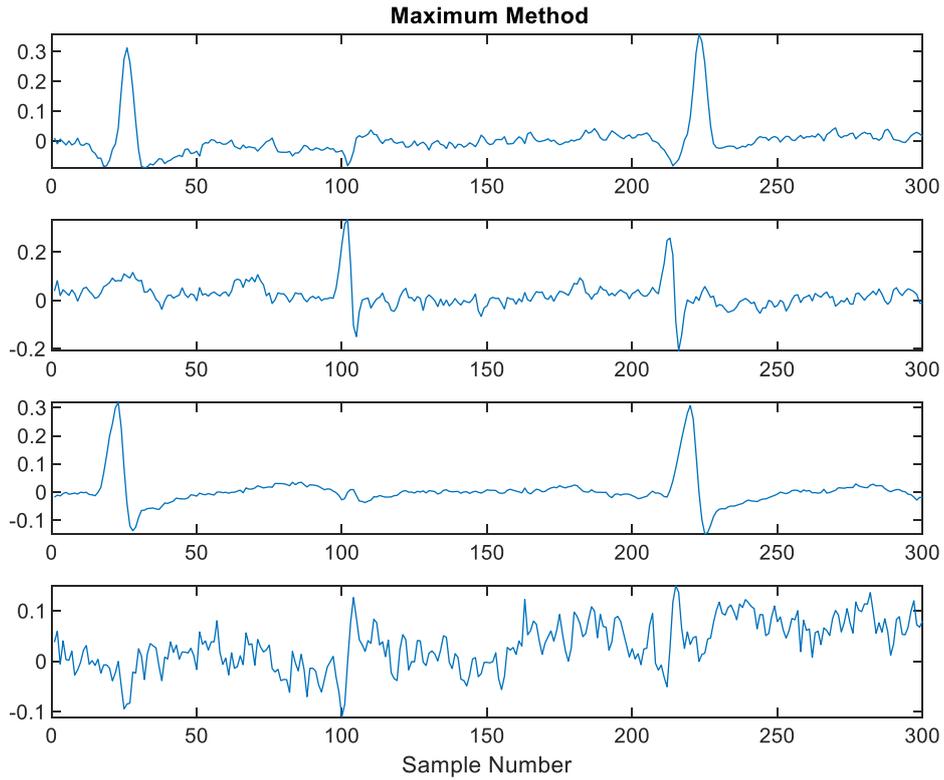

*Figure 41 – Estimated Sources using the Maximum method with whitening*

It can be seen that there is a clear maternal component in the third estimated signal and the fetal contribution can clearly be seen in the second estimated source. The quality of the maternal and fetal estimates are comparable with those obtained using the PCA, although there is observed some maternal contamination of the fetal estimate.

If FastICA is applied with pow3 nonlinearity and using deflation to the data in Figure 39, then the estimated sources are as in Figure 42 below:

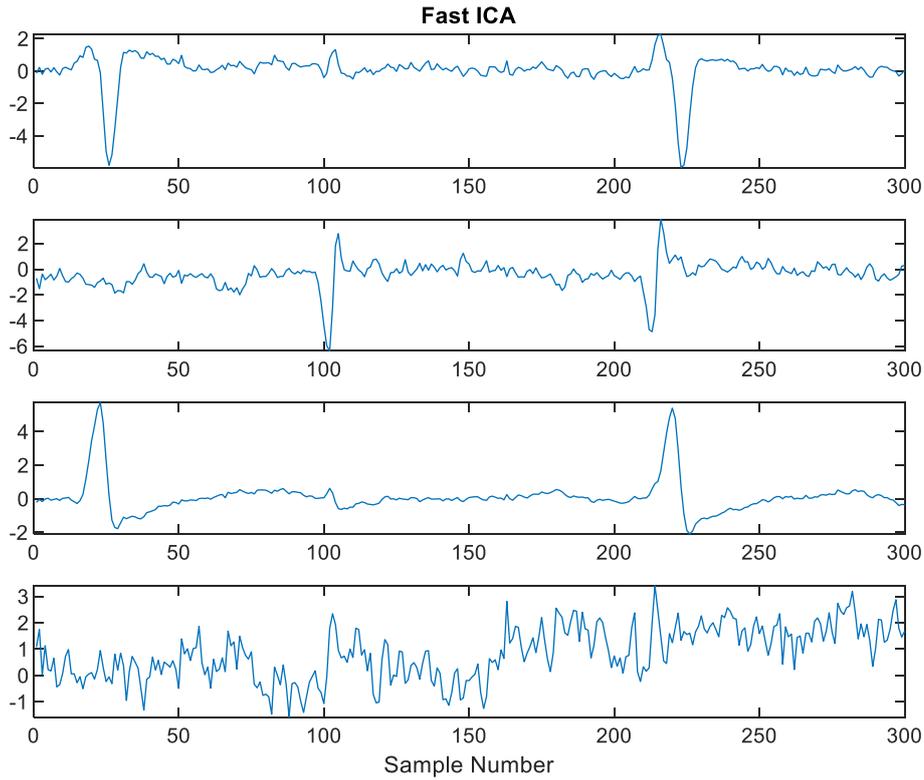

*Figure 42 – Estimated Sources using FastICA*

Comparing Figures 40,,41 and 42 it can be seen that there are no significant differences between the results of PCA, Maximum method and FastICA.

### 4.4.3   4-Lead Physionet Data [31,32]

As a final example, let us look at a set of data taken from the Physionet database [31,32]. The first 3000 samples are taken from the dataset r01.edf.

The original data are shown in Figure 43 below:

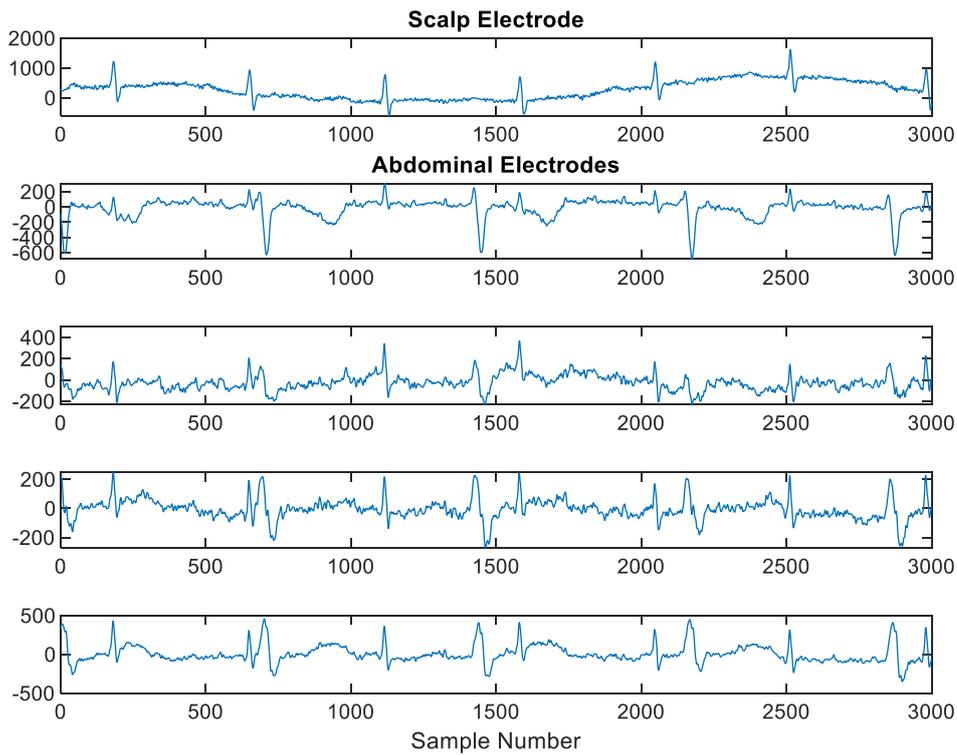

*Figure 43 – First 3000 samples of r01.edf from [31]. The top figure is from a scalp electrode (as a reference), the remaining 4 leads are from the abdomen.*

The scalp electrode lead data at the top of Figure 43 is used as a reference to confirm where the fetal peaks occur. The data that are processed are the four abdominal leads; it can be seen that each lead consists of a mixture of maternal and fetal components.

Firstly, let us look at the results of applying PCA to the abdominal leads, Figure 44:

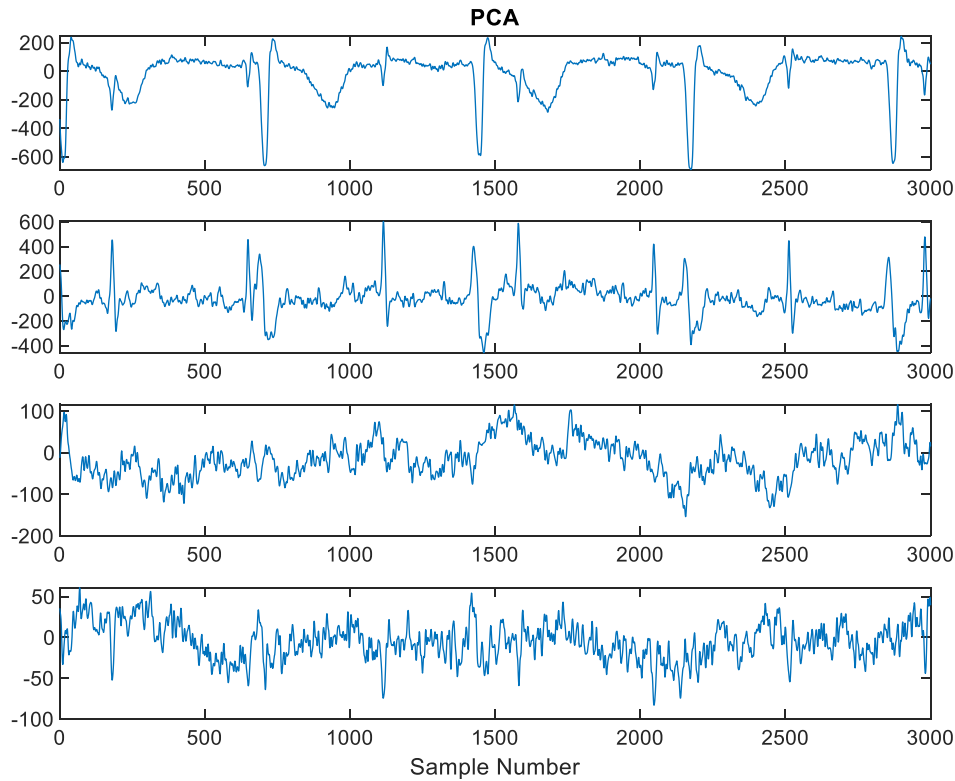

*Figure 44 – Estimated Sources using PCA*

There is no satisfactory separation of fetal and maternal contributions. The principal component contains mainly a maternal contribution, but there is a noticeable fetal component as well. The second component contains comparable maternal and fetal contributions, with there being no useful information in the last two components.

If we now apply the Maximum method with pre-whitening then the outputs are as shown in Figure 45 below:

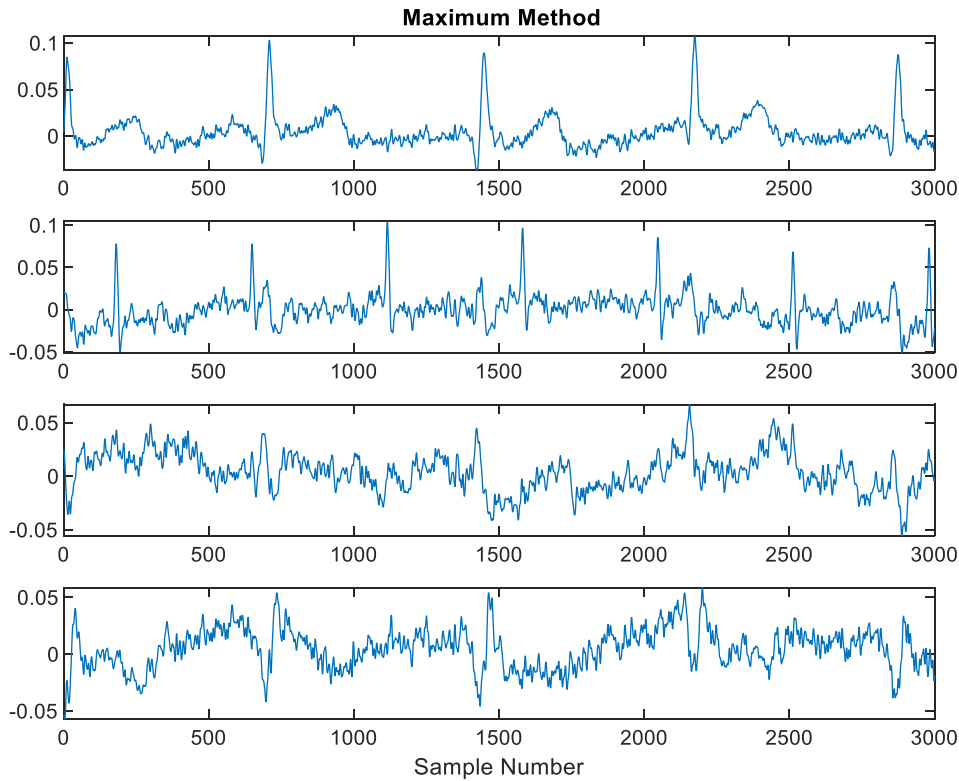

*Figure 45 – Estimated Sources using the Maximum method with whitening*

The resulting estimated sources are significantly better than for PCA. The first component found is mainly maternal with very little fetal contribution. The second estimated source is dominated by the fetal component although contamination by the maternal component can be seen.

The third and fourth components do not carry any useful information.

If we apply FastICA to this dataset then the estimated sources are as shown in Figure 46.

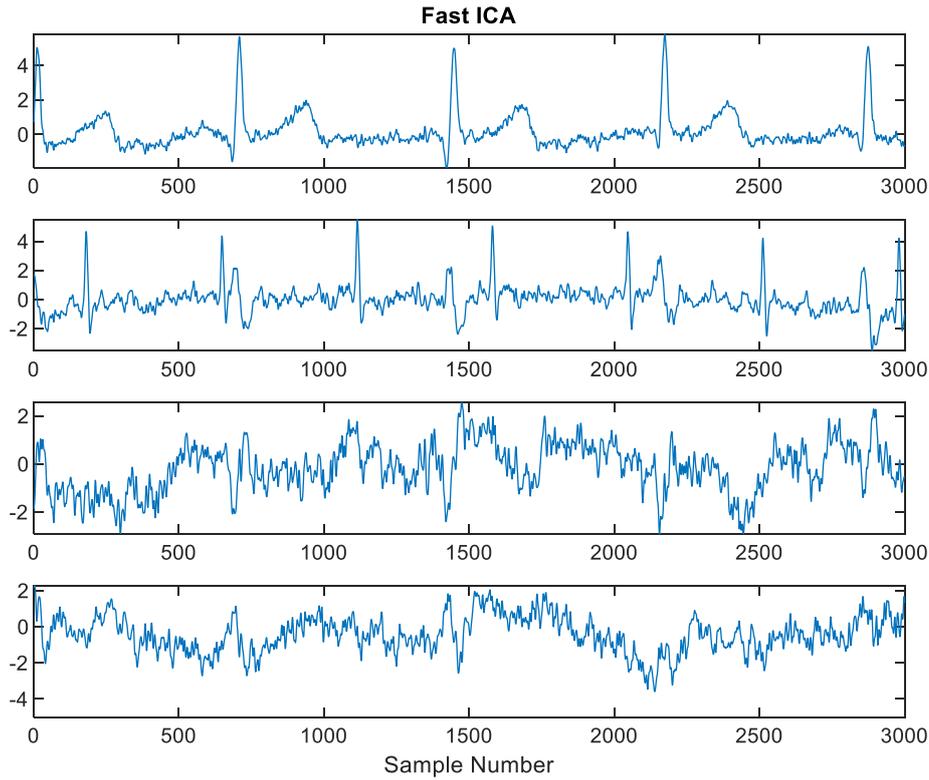

*Figure 46 – Estimated Sources using FastICA*

The results are comparable to those found using the Maximum method with pre-whitening: the second estimated source is mainly fetal with the first estimated source corresponding to the maternal.

To summarise, for this dataset, the Maximum method, with pre-whitening, is better than PCA when estimating fetal and maternal contributions. In addition, the results with the Maximum method are comparable to those obtained using FastICA.

## 5.    Summary of Results

If no whitening is applied to the data then the results when using the Maximum method are very similar to those obtained using the PCA; an example is shown in Figures 34 and 35. This demonstrates that for sparse data with peaks, the Maximum method is doing much the same job as PCA in finding significant directions in phase space.

As expected, pre-whitening is required when using the Maximum method to extract sources that are uncorrelated as this ensures that the directions in phase space corresponding to these sources are orthogonal.

The robustness to noise for the Maximum and PCA methods are compared in Figures 14, 16, 18 and 20, where it is shown that, up to a certain level of noise, the performances of the two methods are comparable. Above a certain noise level though PCA has better performance as this method processes all the data statistically whilst the Maximum method is using information from just one data point in phase space and is hence more susceptible to noise.

In Section 4.3, we saw that in the case where the peaks of two different sources are coincident it is found, as expected, that the Maximum method breaks down as it locates this coincident point in phase space as the principal direction; this is less of a problem with PCA as it is processing all the data.

Data has been analysed from electrodes placed on an expectant mother; data from two different databases have been analysed. Analysis of these data shows comparable results between PCA and the Maximum method for the most part, with Maximum performing better for the Physionet data [31] – however, this is not a general observation for the Maximum method, just for that particular set of data. In general, there is found to be comparable performances for all methods that have been applied to the ECG data that have been analysed in this paper.

## 6. Discussion

In this paper, we have considered the data in *N*-dimensional phase space and chosen the direction corresponding to the most prominent source as the point on the trajectory that is furthest from the origin. Deflation is used to estimate the other sources. There are, however, other related methods of phase space analysis to determine sparse sources in data mixtures and a review is given in [25]. These methods are based on the fact that individual sources can be represented by straight lines in phase space; this can be seen, for example, in Figure 26 where the two straight lines in the phase plot correspond to different sources. In these methods, differences between adjacent data points are taken prior to processing. One advantage of this approach over that used in this paper can be seen in Figure 32. The direction of the maximum in phase space is wrong – this corresponds to a coincidence between the peaks of different sources. However, detecting straight lines (1 and 2) in the phase trajectory corresponding to different sources will lead to the correct directions in phase space corresponding to these sources – see [25] for more details.

Now phase space representations of signals have been used extensively for some time to characterise chaotic systems. There has also been considerable work in applying phase space analysis to separating maternal and fetal components of abdominal signals, the approaches being used being different from the one presented in this paper. In [33] the authors look at the analysis of a single lead abdominal signal. Phase plots are used plotting the signal at time point $n$ against a delayed version at time point $n+d$. The delay time is relatively small at 0.02s so that the maternal ECG is well delineated and the fetal ECG is averaged over and is insignificant. Then the resultant plot is subtracted from the original signal to obtain the fetal signal. Then, to reduce noise, an even smaller delay, $d$, is used so that so that just noise is extracted and then subtracted from output of the previous method to obtain an enhanced fetal signal. In [34], the authors look at single-lead abdominal data and optimise the embedding dimension, delay and clustering threshold to extract the fetal component from abdominal data. As in Reference [33], this is a two-stage process – firstly the maternal contribution is extracted and then this is subtracted from the original data to obtain an estimate of the fetal component in noise and then further processed, as in Schreiber, to reduce noise further on the estimate of the fetal signal. There is varying success dependent on week of pregnancy.

In [35], the authors analyse the data from the Daisy database that was looked at in Sections 4.4.1 and 4.4.2. The authors apply ICA with time-delayed decorrelation before applying Schreiber's method of noise reduction using phase plane plots.

Phase plots are also used to detect QRS complexes in ECG signals, for example in [36]. In this reference the authors propose plotting the signal against its derivative (first difference), which can also be used to locate R-waves; the advantage of this method over those described in [33-35] is that it does not involve finding an optimal delay, $d$.

Phase plots are also used in Vector Cardiography, although in this application special arrangements of leads on the body need to be set up e.g. [37,38]:

Finally, it should also be noted that various BSS techniques have been applied to the separation of sparse sources from signal mixtures, e.g. see [39]. These have involved a statistical analysis of the data trying to find a linear combination of data inputs that, for example, maximises the negentropy [7] or that minimises the fourth order cross-cumulants [6]. The method described in this paper uses essentially one data point to determine each source, the data point corresponding to a maximum in phase space. It can be seen that, although this is a relatively simple approach, in many cases the quality of the estimated sources are comparable with those obtained from methods that are more sophisticated as long as there is not a coincidence of major peaks between the two sources.  The reason for this good performance is that the sources are sparse and have peaks that occur in specific time segments where the contributions from the other sources are negligible.  The disadvantage of the Maximum method is that it does not work well for separating non-sparse sources, whilst many BSS methods are able to deal with such sources, as these methods do not assume sparsity of sources.

## 7. Conclusions

A simple method of phase space analysis has been presented to separate sparse sources from mixtures.  This involves whitening the data first (using, for example, the Gram Schmidt method and normalisation) then looking for the direction of the maximum of the phase plot.  Source estimates are subtracted out using the deflation approach to estimate the other sources.  The method has been applied to both simulated signals and actual ECG signals taken from an expectant mother.  In general, the performance of this method is comparable to the PCA and Fast ICA.  However, the method breaks down, as expected, if there is a coincidence of sources because it just uses one point in phase space that could correspond to this coincidence.  The method relies on the direction in phase space being from one source only; better performances in this case can be obtained from the methods described in [17],[18] and [25]..

As an alternative to the Gram-Schmidt method, if one applies PCA to some data and then normalise the components, then these estimates can be used as the whitened components, to which the Maximum method can be applied.  Therefore we can also think of the Maximum method as a post-processing method for the PCA.